\pdfoutput=1
\documentclass[12pt,a4paper]{article}
\usepackage{ifthen} 
\newboolean{pdflatex}
\setboolean{pdflatex}{true} 

\newboolean{articletitles}
\setboolean{articletitles}{true} 

\newboolean{uprightparticles}
\setboolean{uprightparticles}{false} 

\def\paperauthors{LHCb collaboration} 
\def\paperasciititle{Amplitude Analysis of Bm2DstarmDpKm} 
\def\papertitle{
Observation of new charmonium(-like) states in $B^+ \to D^{*\pm} D^{\mp} K^+$ decays}
\def\paperkeywords{{High Energy Physics}, {LHCb}} 
\def\papercopyright{\the\year\ CERN for the benefit of the LHCb collaboration} 
\def\paperlicence{CC BY 4.0 licence}
\def\paperlicenceurl{https://creativecommons.org/licenses/by/4.0/}

\usepackage[top=1in, bottom=1.25in, left=1in, right=1in]{geometry}

\usepackage{microtype}
\usepackage{lineno}  
\usepackage{xspace} 
\usepackage{caption} 

\usepackage{graphicx}  
\usepackage{color}
\usepackage{colortbl}
\graphicspath{{./figs/}} 

\usepackage{amsmath} 
\usepackage{amssymb}
\usepackage{amsfonts}
\usepackage{upgreek} 

\newcommand*\patchAmsMathEnvironmentForLineno[1]{%
\expandafter\let\csname old#1\expandafter\endcsname\csname #1\endcsname
\expandafter\let\csname oldend#1\expandafter\endcsname\csname
end#1\endcsname
 \renewenvironment{#1}%
   {\linenomath\csname old#1\endcsname}%
   {\csname oldend#1\endcsname\endlinenomath}%
}
\newcommand*\patchBothAmsMathEnvironmentsForLineno[1]{%
  \patchAmsMathEnvironmentForLineno{#1}%
  \patchAmsMathEnvironmentForLineno{#1*}%
}
\AtBeginDocument{%
\patchBothAmsMathEnvironmentsForLineno{equation}%
\patchBothAmsMathEnvironmentsForLineno{align}%
\patchBothAmsMathEnvironmentsForLineno{flalign}%
\patchBothAmsMathEnvironmentsForLineno{alignat}%
\patchBothAmsMathEnvironmentsForLineno{gather}%
\patchBothAmsMathEnvironmentsForLineno{multline}%
\patchBothAmsMathEnvironmentsForLineno{eqnarray}%
}

\usepackage{hyperxmp}

\usepackage[pdftex,
            pdfauthor={\paperauthors},
            pdftitle={\paperasciititle},
            pdfkeywords={\paperkeywords},
            pdfcopyright={Copyright (C) \papercopyright},
            pdflicenseurl={\paperlicenceurl}]{hyperref}

\usepackage[bottom,flushmargin,hang,multiple]{footmisc}

\usepackage[all]{hypcap} 

\usepackage{xspace} 
\usepackage{upgreek}

\def\lhcb   {\mbox{LHCb}\xspace}

\def\MagUp {\mbox{\em Mag\kern -0.05em Up}\xspace}

\ifthenelse{\boolean{uprightparticles}}%
{

 \def\Ppi         {\ensuremath{\uppi}\xspace}

 \def\PDelta      {\ensuremath{\Delta}\xspace}                 
 \def\PXi         {\ensuremath{\Xi}\xspace}                 
 \def\PLambda     {\ensuremath{\Lambda}\xspace}                 
 \def\PSigma      {\ensuremath{\Sigma}\xspace}                 
 \def\POmega      {\ensuremath{\Omega}\xspace}                 
 \def\PUpsilon    {\ensuremath{\Upsilon}\xspace}
 \let\oldPi\Pi
 \def\PPi         {\ensuremath{\oldPi}\xspace}

 \def\PB      {\ensuremath{\mathrm{B}}\xspace}                 
                  
 \def\PD      {\ensuremath{\mathrm{D}}\xspace}

 \def\PK      {\ensuremath{\mathrm{K}}\xspace}

 \def\Pi      {\ensuremath{\mathrm{i}}\xspace}

 \def\Ps      {\ensuremath{\mathrm{s}}\xspace}

 \def\thebaroffset{0.0em}
}
{

 \def\Ppi         {\ensuremath{\pi}\xspace}

 \mathchardef\PDelta="7101
 \mathchardef\PXi="7104
 \mathchardef\PLambda="7103
 \mathchardef\PSigma="7106
 \mathchardef\POmega="710A
 \mathchardef\PUpsilon="7107
 \mathchardef\PPi="7105
                  
 \def\PB      {\ensuremath{B}\xspace}                 
                  
 \def\PD      {\ensuremath{D}\xspace}

 \def\PK      {\ensuremath{K}\xspace}

 \def\Pi      {\ensuremath{i}\xspace}

 \def\Ps      {\ensuremath{s}\xspace}

 \def\thebaroffset{0.18em}
}
\newcommand{\offsetoverline}[2][\thebaroffset]{\kern #1\overline{\kern -#1 #2}}%

\makeatletter
\ifcase \@ptsize \relax
  \newcommand{\miniscule}{\@setfontsize\miniscule{4}{5}}
\or
  \newcommand{\miniscule}{\@setfontsize\miniscule{5}{6}}
\or
  \newcommand{\miniscule}{\@setfontsize\miniscule{5}{6}}
\fi
\makeatother

\DeclareRobustCommand{\optbar}[1]{\shortstack{{\miniscule (\rule[.5ex]{1.25em}{.18mm})}
  \\ [-.7ex] $#1$}}

\def\squark    {{\ensuremath{\Ps}}\xspace}

\def\pion   {{\ensuremath{\Ppi}}\xspace}

\def\pip    {{\ensuremath{\pion^+}}\xspace}
\def\pim    {{\ensuremath{\pion^-}}\xspace}
\def\pipm   {{\ensuremath{\pion^\pm}}\xspace}

\def\kaon    {{\ensuremath{\PK}}\xspace}

\def\KorKbar {\kern \thebaroffset\optbar{\kern -\thebaroffset \PK}{}\xspace}

\def\Kp      {{\ensuremath{\kaon^+}}\xspace}

\def\Kpm     {{\ensuremath{\kaon^\pm}}\xspace}

\def\Dbar    {{\ensuremath{\offsetoverline{\PD}}}\xspace}
\def\D       {{\ensuremath{\PD}}\xspace}

\def\DorDbar {\kern \thebaroffset\optbar{\kern -\thebaroffset \PD}\xspace}
\def\Dz      {{\ensuremath{\D^0}}\xspace}
\def\Dzb     {{\ensuremath{\Dbar{}^0}}\xspace}
\def\Dp      {{\ensuremath{\D^+}}\xspace}
\def\Dm      {{\ensuremath{\D^-}}\xspace}
\def\Dpm     {{\ensuremath{\D^\pm}}\xspace}
\def\Dmp     {{\ensuremath{\D^\mp}}\xspace}
\def\DpDm    {\ensuremath{\Dp {\kern -0.16em \Dm}}\xspace}
\def\Dstar   {{\ensuremath{\D^*}}\xspace}

\def\Dstarz  {{\ensuremath{\D^{*0}}}\xspace}
\def\Dstarzb {{\ensuremath{\Dbar{}^{*0}}}\xspace}

\def\Dstarp  {{\ensuremath{\D^{*+}}}\xspace}
\def\Dstarm  {{\ensuremath{\D^{*-}}}\xspace}
\def\Dstarpm {{\ensuremath{\D^{*\pm}}}\xspace}
\def\Dstarmp {{\ensuremath{\D^{*\mp}}}\xspace}

\def\Dsp     {{\ensuremath{\D^+_\squark}}\xspace}

\def\B       {{\ensuremath{\PB}}\xspace}

\def\BorBbar {\kern \thebaroffset\optbar{\kern -\thebaroffset \PB}\xspace}
\def\Bz      {{\ensuremath{\B^0}}\xspace}

\def\Bd      {{\ensuremath{\B^0}}\xspace}

\def\BdorBdbar {\kern \thebaroffset\optbar{\kern -\thebaroffset \Bd}\xspace}
\def\Bu      {{\ensuremath{\B^+}}\xspace}

\def\Bp      {{\ensuremath{\Bu}}\xspace}

\def\Bs      {{\ensuremath{\B^0_\squark}}\xspace}

\def\BsorBsbar {\kern \thebaroffset\optbar{\kern -\thebaroffset \Bs}\xspace}

\def\Y#1S{\ensuremath{\PUpsilon{(#1S)}}\xspace}

\def\LorLbar     {\kern \thebaroffset\optbar{\kern -\thebaroffset \PLambda}\xspace}

\newcommand{\decay}[2]{\ensuremath{#1\!\to #2}\xspace} 

\def\to                 {\ensuremath{\rightarrow}\xspace}

\def\AT#1     {\ensuremath{A_{\mathrm{T}}^{#1}}\xspace}           

\def\C#1      {\ensuremath{\mathcal{C}_{#1}}\xspace}                       
\def\Cp#1     {\ensuremath{\mathcal{C}_{#1}^{'}}\xspace}                    
\def\Ceff#1   {\ensuremath{\mathcal{C}_{#1}^{\mathrm{(eff)}}}\xspace}        
\def\Cpeff#1  {\ensuremath{\mathcal{C}_{#1}^{'\mathrm{(eff)}}}\xspace}       
\def\Ope#1    {\ensuremath{\mathcal{O}_{#1}}\xspace}                       
\def\Opep#1   {\ensuremath{\mathcal{O}_{#1}^{'}}\xspace}                    

\newcommand{\aunit}[1]{\ensuremath{\text{\,#1}}}       

\newcommand{\tev}{\aunit{Te\kern -0.1em V}\xspace}
\newcommand{\gev}{\aunit{Ge\kern -0.1em V}\xspace}
\newcommand{\mev}{\aunit{Me\kern -0.1em V}\xspace}
\newcommand{\kev}{\aunit{ke\kern -0.1em V}\xspace}
\newcommand{\ev}{\aunit{e\kern -0.1em V}\xspace}
 
\newcommand{\mevc}{\ensuremath{\aunit{Me\kern -0.1em V\!/}c}\xspace}
\newcommand{\gevc}{\ensuremath{\aunit{Ge\kern -0.1em V\!/}c}\xspace}
\newcommand{\mevcc}{\ensuremath{\aunit{Me\kern -0.1em V\!/}c^2}\xspace}
\newcommand{\gevcc}{\ensuremath{\aunit{Ge\kern -0.1em V\!/}c^2}\xspace}

\def\fb   {\ensuremath{\aunit{fb}}\xspace}
\def\invfb   {\ensuremath{\fb^{-1}}\xspace}

\newcommand{\chisq}{\ensuremath{\chi^2}\xspace}

\def\gsim{{~\raise.15em\hbox{$>$}\kern-.85em
          \lower.35em\hbox{$\sim$}~}\xspace}
\def\lsim{{~\raise.15em\hbox{$<$}\kern-.85em
          \lower.35em\hbox{$\sim$}~}\xspace}

\def\evtgen     {\mbox{\textsc{EvtGen}}\xspace}

\def\geant      {\mbox{\textsc{Geant4}}\xspace}

\def\photos     {\mbox{\textsc{Photos}}\xspace}

\def\pythia     {\mbox{\textsc{Pythia}}\xspace}

\def\tell1  {TELL1\xspace}
\def\ukl1   {UKL1\xspace}

\newcommand{\ie}{\mbox{\itshape i.e.}\xspace}

\newcommand{\lhcborcid}[1]{\href{https://orcid.org/#1}{\hspace*{0.1em}\raisebox{-0.45ex}{\includegraphics[width=1em]{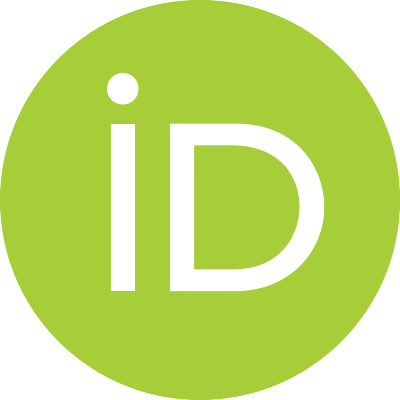}}}}

\usepackage{cite} 
\usepackage{mciteplus}
\usepackage{makecell}
\usepackage{multirow}

\usepackage{longtable} 

\begin{document}

\renewcommand{\thefootnote}{\fnsymbol{footnote}}
\setcounter{footnote}{1}


\begin{titlepage}
\pagenumbering{roman}

\vspace*{-1.5cm}
\centerline{\large EUROPEAN ORGANIZATION FOR NUCLEAR RESEARCH (CERN)}
\vspace*{1.5cm}
\noindent
\begin{tabular*}{\linewidth}{lc@{\extracolsep{\fill}}r@{\extracolsep{0pt}}}
\ifthenelse{\boolean{pdflatex}}
{\vspace*{-1.5cm}\mbox{\!\!\!\includegraphics[width=.14\textwidth]{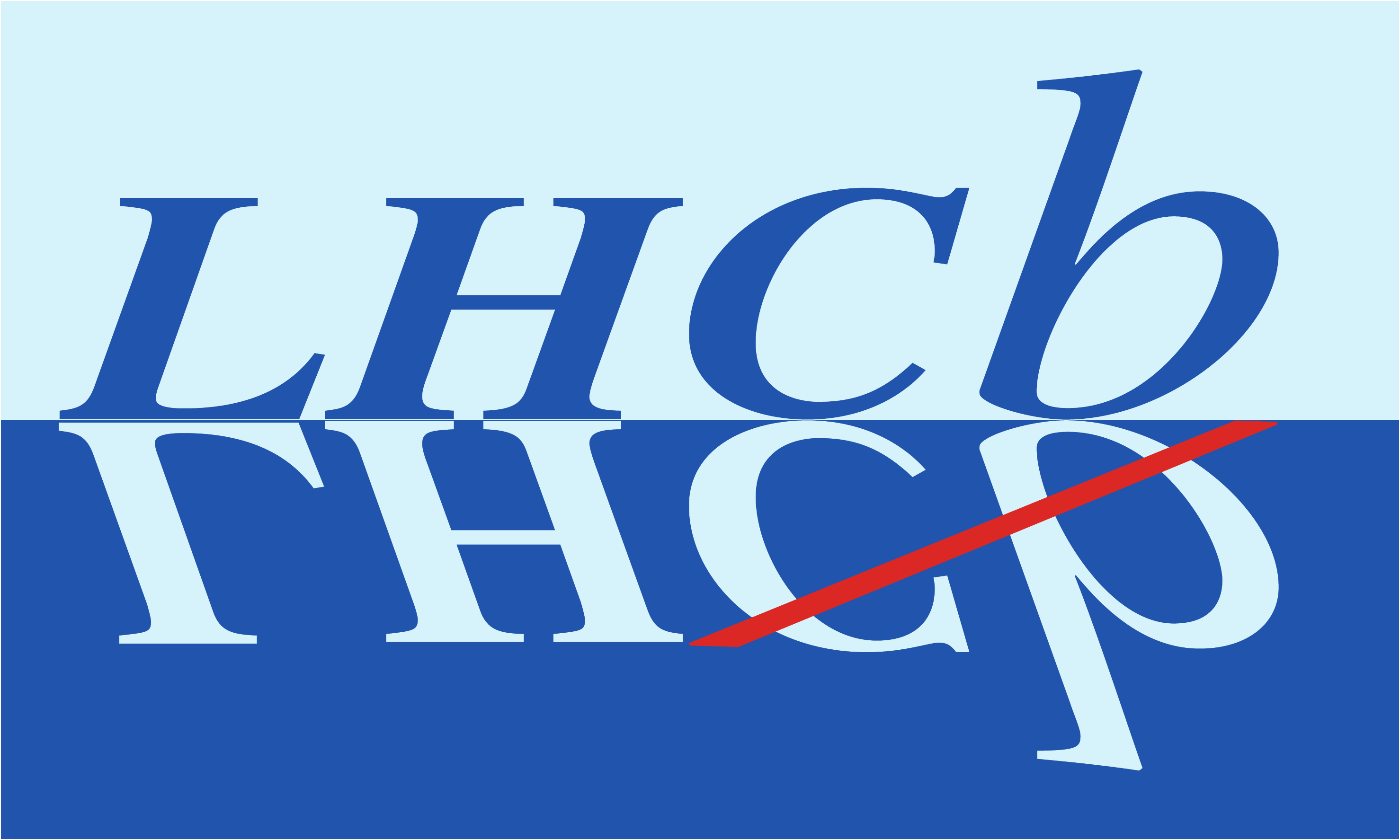}} & &}%
{\vspace*{-1.2cm}\mbox{\!\!\!\includegraphics[width=.12\textwidth]{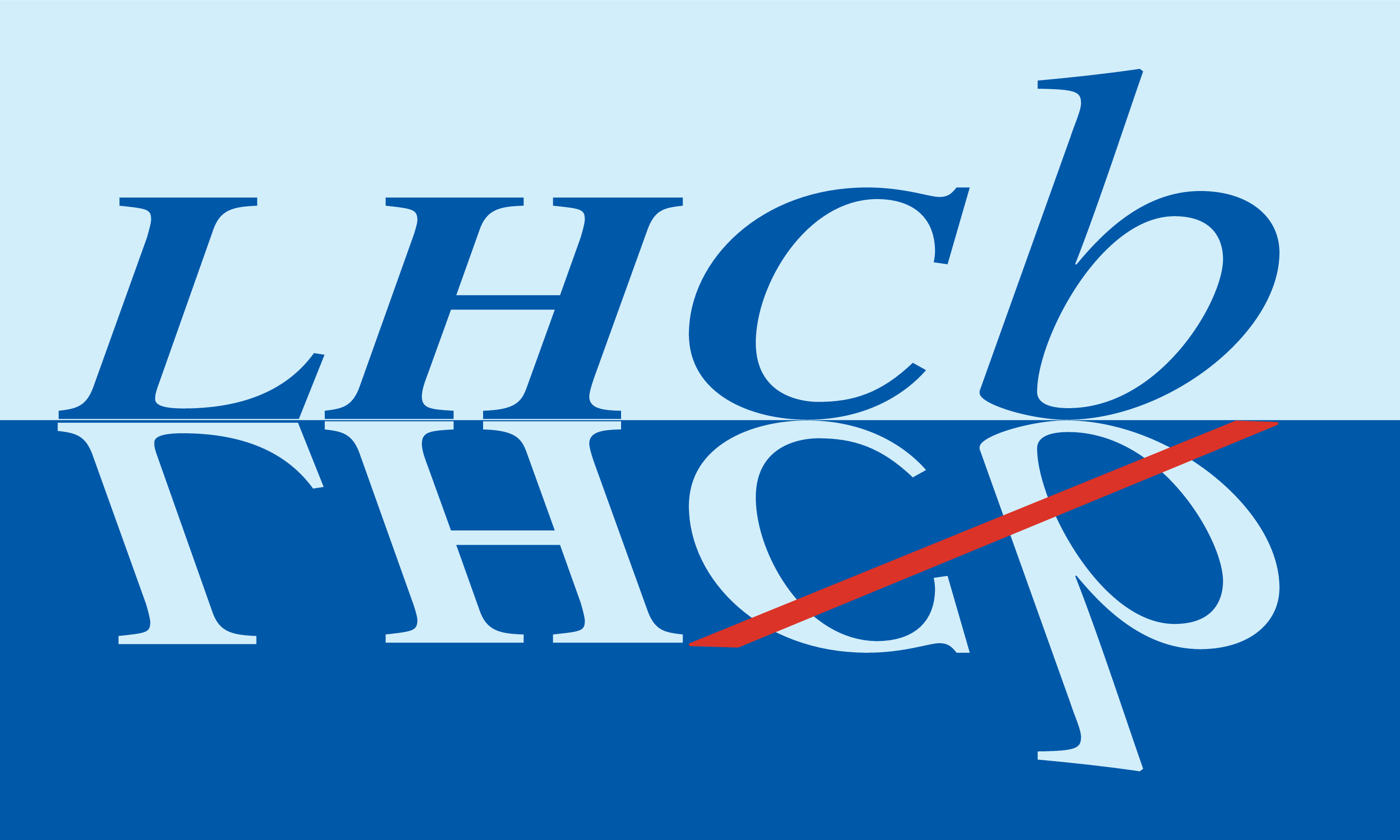}} & &}%
\\
 & & CERN-EP-2024-096 \\  
 & & LHCb-PAPER-2023-047 \\  
 & & 22 Jan 2025 \\
 & & \\
\end{tabular*}

\vspace*{3.0cm}

{\normalfont\bfseries\boldmath\huge
\begin{center}
  \papertitle 
\end{center}
}

\vspace*{2.0cm}

\begin{center}
\paperauthors\footnote{Authors are listed at the end of this paper.}
\end{center}

\vspace{\fill}

\begin{abstract}
  \noindent
  A study of resonant structures in  $B^{+}\rightarrow{D^{\ast+}D^{-}K^{+}}$ and $B^{+}\rightarrow{D^{\ast-}D^{+}K^{+}}$ decays is performed, 
  using proton-proton collision data at centre-of-mass energies of $\sqrt{s}=7, 8$, and $13\tev$ recorded by the LHCb experiment, corresponding to an integrated luminosity of $9\invfb$. 
  A simultaneous amplitude fit is performed to the two channels with contributions from resonances decaying to $D^{\ast-}D^{+}$ and $D^{\ast+}D^{-}$ states linked by $C$ parity.
  This procedure allows the $C$-parities of resonances in the $D^{\ast\pm}D^{\mp}$ mass spectra to be determined. 
  Four charmonium(-like) states are observed decaying into $D^{\ast\pm}D^{\mp}$: $\eta_c(3945)$, $h_c(4000)$, $\chi_{c1}(4010)$ and $h_c(4300)$, with quantum numbers $J^{PC}$ equal to $0^{-+}$,  $1^{+-}$, $1^{++}$ and $1^{+-}$, respectively. At least three of these states have not been observed previously.  
  In addition, the existence of the $T_{\bar{c}\bar{s}0}^{*}(2870)^{0}$ and $T_{\bar{c}\bar{s}1}^{*}(2900)^{0}$ resonances in the $D^-K^+$ mass spectrum, already observed in the $B^+ \to D^+ D^- K^+$ decay, is confirmed in a different production channel.
  
\end{abstract}

\vspace*{1.0cm}

\begin{center}
 Published in 
  Phys.~Rev.~Lett. 
\end{center}

\vspace{\fill}

{\footnotesize 
\centerline{\copyright~\papercopyright. \href{\paperlicenceurl}{\paperlicence}.}}
\vspace*{2mm}

\end{titlepage}


\newpage
\setcounter{page}{2}
\mbox{~}

\renewcommand{\thefootnote}{\arabic{footnote}}
\setcounter{footnote}{0}

\cleardoublepage

\pagestyle{plain} 
\setcounter{page}{1}
\pagenumbering{arabic}

\newcommand{\Xnew}[2][]{
    \ifthenelse{\equal{#2}{X0(3945)}}
        {\ensuremath{\eta_c(3945)}}
    {\ifthenelse{\equal{#2}{X1(4000)}}
        {\ensuremath{h_{c}(4000)}}
    {\ifthenelse{\equal{#2}{X1(4010)}}
        {\ensuremath{\chi_{c1}(4010)}} 
    {\ifthenelse{\equal{#2}{X1(4300)}}
        {\ensuremath{h_c(4300)}}
    {\ifthenelse{\equal{#2}{X(4020)}}
        {\ensuremath{T_{c\bar{c}}(4020)^{0}}}
    {\ifthenelse{\equal{#2}{X0(2900)}}
        {\ensuremath{T_{\bar{c}\bar{s}0}^{*}(2870)^{0}}}
    {\ifthenelse{\equal{#2}{X1(2900)}}
        {\ensuremath{T_{\bar{c}\bar{s}1}^{*}(2900)^{0}}}
    {\ifthenelse{\equal{#2}{X01(2900)}}
        {\ensuremath{T_{\bar{c}\bar{s}0,1}^{*0}}}
    {\ifthenelse{\equal{#2}{Zc(3900)}}
        {\ensuremath{T_{c\bar{c}1}(3900)^{0}}}
    {\ifthenelse{\equal{#2}{Zc(4020)}}
        {\ensuremath{T_{c\bar{c}}(4020)^{0}}}
    {\ifthenelse{\equal{#2}{Tcsbar(2900)++}}
        {\ensuremath{T_{c\bar{s}0}^{*}(2900)^{++}}}
    {\ifthenelse{\equal{#2}{X1(4450)}}
        {\ensuremath{\chi_{c1}(4450)}}
    {\ifthenelse{\equal{#2}{Zc(4430)}}
        {\ensuremath{T_{c\bar{c}1}(4430)^0}}
        {\ensuremath{X(#2)}}
    }}}}}}}}}}}}
}

The observation of two open-charmed resonant structures $\Xnew{X0(2900)}$ and $\Xnew{X1(2900)}$, consistent with tetraquark states with minimal quark content $[\bar{c}\bar{s}ud]$, was reported by the \lhcb collaboration in the \decay{\Bp}{\Dp\Dm\Kp} decay~\cite{LHCb-PAPER-2020-024,LHCb-PAPER-2020-025}.\footnote{The inclusion of charge conjugate processes is implied throughout the Letter, unless otherwise specified. The hadron naming scheme of Ref.~\cite{naming} is used.}
Confirmation of these states in other processes is critical to further establish their existence and improve understanding of their nature. 
Later, a pair of open-charmed exotic hadrons, one doubly charged and one neutral, were observed decaying to $\Dsp\pipm$ in the \decay{\Bp}{\Dm \Dsp\pip} and \decay{\Bz}{\Dzb \Dsp\pim} processes, with quark contents $[c\bar{s}u\bar{d}]$ and $[c\bar{s}\bar{u}d]$~\cite{LHCb-PAPER-2022-026, LHCb-PAPER-2022-027}, respectively. 
The $\Bp\rightarrow{\Dstarp\Dm\Kp}$ and $\Bp\rightarrow{\Dstarm\Dp\Kp}$ decays are clean processes to search for open-charmed exotic states, as conventional excited $\Dsp$ mesons can not be formed in them. 
In addition, these decays are useful for investigation of \mbox{charmonium(-like)} states decaying into $\Dstarpm\Dmp$. 
Among the observed \mbox{charmonium(-like)} states, more than a dozen cannot be interpreted as $c\bar{c}$ charmonium states in the quark model~\cite{PDG2022,Brambilla:2019esw,Olsen:2017bmm}.
However, few measurements have been made of \mbox{charmonium(-like)} states decaying to $\Dstarpm\Dmp$.
In the studies which have been performed,  the $X(3940)$ state was seen in \mbox{$e^+ e^- \to J/\psi  D^{*0/+}\Dbar{}^{0/-}$}~\cite{Belle:2005lik,Belle:2007woe}, and the $\Xnew{Zc(3900)}$ state was seen in $e^+e^-\to \pi^0 D^{*0/+}\Dbar{}^{0/-}$~\cite{BESIII:2015ntl} processes, both of which exploit $e^+e^-$ annihilations and therefore cannot be used to explore the full possible range of $J^{PC}$ quantum numbers.
Studies in $B$ decays are of special interest as the different production mechanism to that of $e^+ e^-$ collisions may affect the rates of  \mbox{charmonium(-like)} states, providing 
opportunities to both understand the nature of already observed states and to uncover new ones.

In this Letter, the results of a simultaneous analysis of the $\Bp\rightarrow{\Dstarp\Dm\Kp}$ and \mbox{$\Bp\rightarrow{\Dstarm\Dp\Kp}$} decays are presented.
The analysis is based on proton-proton ($pp$) collision data collected by the \lhcb experiment, corresponding to an integrated luminosity of $9 \invfb$.  
Due to $C$-parity conservation in strong decays, any single \mbox{charmonium(-like)} resonance $R$ must contribute equally to the $\Bp \to R(\Dstarp\Dm) \Kp$ and $\Bp \to R(\Dstarm\Dp) \Kp$ processes, but interference between resonances with different $C$-parity can result in differences between the two final states.
By linking the decay amplitudes for $\Bp \to R(\Dstarp\Dm) \Kp$ and $\Bp \to R(\Dstarm\Dp) \Kp$ by $C$-parity, one can determine the $C$-parities of the resonances. 
This method is implemented for the first time in the analysis reported here.

The \lhcb detector is a single-arm forward spectrometer covering the pseudorapidity range $2 < \eta < 5$, described in detail in Refs.~\cite{LHCb-DP-2008-001,LHCb-DP-2014-002}. It is designed specifically for the study of particles containing $b$ or $c$ quarks. Simulation is required to model the effects of the detector acceptance and the imposed selection requirements. 
In the simulation, $pp$ collisions are generated using \pythia~\cite{Sjostrand:2007gs,*Sjostrand:2006za} with a specific \lhcb configuration~\cite{LHCb-PROC-2010-056}. 
Decays of unstable particles are described by \evtgen~\cite{Lange:2001uf}, in which final-state radiation is generated using \photos~\cite{davidson2015photos}. 
The interaction of the generated particles with the detector and its response are implemented using the \geant toolkit~\cite{Allison:2006ve, *Agostinelli:2002hh} as described in Ref.~\cite{LHCb-PROC-2011-006}.

The same selection criteria as in a previous measurement of the branching fractions of $\Bp\rightarrow{\Dstarp\Dm\Kp}$ and $\Bp\rightarrow{\Dstarm\Dp\Kp}$ decays~\cite{LHCb-PAPER-2020-006} are used.
Charmed mesons are reconstructed using the $\Dstarm \to \Dzb \pi^-$, $\Dzb\to\Kp\pim$, $\Dzb\to\Kp\pim\pim\pip$ and \mbox{$\Dm\to\Kp\pim\pim$} decays. 
Preselections are applied  to identify well-reconstructed tracks displaced from primary vertices (PVs) of $pp$ collisions and with large transverse momenta. 
Charmed meson candidates are required to have reconstructed masses consistent with the known values~\cite{PDG2022} and to have high-quality reconstructed  vertices. 
Requirements are also applied on the quality of the $B$-candidate decay vertex and its displacement from any PV. 
A multivariate classifier, based on a boosted decision tree (BDT)~\cite{Breiman,AdaBoost} algorithm in the TMVA toolkit~\cite{Hocker:2007ht,*TMVA4}, is employed to further reduce combinatorial backgrounds using topological and particle identification (PID) information.  
In addition, backgrounds with one or zero charmed mesons in the final states are reduced by requiring the reconstructed $D^{0/+}$ candidates to have a significant flight distance from the \Bp decay vertex. 
The selected sample is divided into subsamples corresponding to two LHC run periods and different $\Dzb$ and $B^+$ decay modes.
For each subsample, a fit is performed to the  $M(\Dstarpm\Dmp\Kp)$ distribution in the mass range $[5210, 5390]\mev$ to estimate the background fraction,  $f_{\rm bg}$, in the signal region defined as $M(\Dstarpm\Dmp\Kp) \in [5260, 5300] \mev$, as done in the previous branching fraction measurement~\cite{LHCb-PAPER-2020-006}.\footnote{Natural units with $\hbar = c = 1$ are used throughout the Letter.}  
In total,  $1636 \pm 43$ and $1772\pm 44$ signal decays are found  in the signal region for the $\Bp\rightarrow{\Dstarp\Dm\Kp}$ and $\Bp\rightarrow{\Dstarm\Dp\Kp}$ decay modes, respectively, both with a background fraction of around $5\%$.

A subsequent amplitude fit, using an unbinned maximum-likelihood method, is performed simultaneously on the four sub samples.
The log-likelihood (LL) for each subsample is defined as
\begin{equation}\label{eq:nll}
\ln L = \sum_{i} \ln \left[(1-f_{\rm bg}) \frac{|\mathcal{A}(x_i)|^2}{\int |\mathcal{A}(x)|^2\epsilon(x) \mathrm{d} \Phi(x)} + f_{\rm bg}\frac{B(x_i)}{\epsilon(x_i)}   \right],
\end{equation}
where $x_i$ is a point in the phase space ($\Phi$) for candidate $i$ of the considered decay.  
The normalization factors are calculated using simulated samples~\cite{Garmash_2005}, from which the efficiency $\epsilon(x)$ is obtained by applying the same criteria as  used for data, and correcting for  
data-simulation differences in the PID~\cite{LHCb-PUB-2016-021}, tracking~\cite{LHCb-TDR-015} and trigger response~\cite{LHCb-PUB-2014-039}. 
The background fraction 
$f_{\rm bg}$ 
is fixed to the value from the preceding mass fit.
The function $B(x)$ is the background probability density function (PDF). 
The background PDF is modeled using candidates in the $B$-candidate mass sideband regions, $M(\Dstarpm\Dmp\Kp) \in [5220, 5240]$ or $[5320, 5340]\mev$.

The amplitude $\mathcal{A}(x)$ is the coherent sum of all resonant and nonresonant (NR) contributions, 
denoted with subscripts $j,k,l$ for resonances decaying into different final states,
\begin{eqnarray}\label{eq:amplitude}
\mathcal{A}(x) &=& \frac{1 + d}{2} \left\{ 
    \sum_{j \in R(\Dstarpm\Dmp)} c_j A_j(x)+ 
    \sum_{k \in R(\Dstarm\Kp,~\Dp\Kp)}c_k A_k(x) \right\} \\\nonumber
&+&  \frac{1 - d}{2}  \left\{ 
    \sum_{j \in R(\Dstarpm\Dmp)} C_j\times c_j A_j(x) + 
    \sum_{l \in R(\Dstarp\Kp,~\Dm\Kp)}c_lA_l(x) \right\},
\end{eqnarray}
where $d = +1$ for $\Bp\rightarrow{\Dstarm\Dp\Kp}$ decays and $d=-1$ for $\Bp\rightarrow{\Dstarp\Dm\Kp}$ decays.
The $C$-parity of the resonance $R$ is denoted as $C_j$. The branching fractions of $B^{+}\rightarrow RK^{+}$ decays followed by $R\rightarrow \Dstarp \Dm$ and $R\rightarrow \Dstarm \Dp$ states are equal, guaranteed by \mbox{$C$-parity} conservation in strong decays of the resonance $R$.
No such constraints are applied on the amplitudes for resonances decaying into $\Dstarmp\Kp$ or $\Dpm\Kp$ states. 
The complex coefficients $c_{j/k/l}$ are determined from the fit to data, with one of them fixed to unity as a reference.
The amplitude of each resonance, $A_{j/k/l}(x)$, is constructed using the helicity formalism. 

Both $S$-wave and $D$-wave amplitudes contribute significantly to the decays \mbox{$R \to \Dstarpm\Dmp$} when $R$ has spin-parity $J^P = 1^+$. 
The line shapes for these partial waves are described by 
\begin{equation} \label{eq:SDRBW}
f_{R,S/D}(m) = \frac{\gamma_{S/D}}{m_0^2 - m^2 - i m_0 [\gamma_{S}^2 \Gamma_{\!S}(m)  + \gamma_{D}^2 \Gamma_{\!D}(m)]}\,,
\end{equation}
where $\gamma_S$ and $\gamma_D$ denote $S$- and $D$-wave coupling constants determined from the fit, with normalization condition $\gamma_S^2 + \gamma_D^2 =1$. 
The mass-dependent width is \mbox{$\Gamma(m) = \Gamma_0 (m_0/m)\left(q/q_0\right)^{2l+1} B_{l}'^2(q,q_0,d)$}, where $l$ corresponds to the angular momentum between the two decay products of the resonance $R$, $B'_{l}(q,q_0,d)$ is the Blatt--Weisskopf barrier factor~\cite{Blatt:1952ije} with $d=3.0\gev^{-1}$, 
$q$ ($q_0$) denotes the momentum of the decay products in the rest frame of the resonance at the reconstructed  mass $m$ (pole mass $m_0$) and $\Gamma_0$ is the width of the resonance. 
For all other resonances, the Breit--Wigner function $f_{R}(m) = 1/[m_0^2 - m^2 - i m_0\Gamma(m)]$ is used. 
An \textit{ad-hoc} formula~\cite{Back:2017zqt} is used to calculate $q_0$ if $m_0$ is smaller than the $M(\Dstarpm \Dmp)$ threshold.

A large range of possible models are fitted to  data, with the LL value used as the primary measure of relative goodness-of-fit, assisted by binned $\chisq$ tests.
The final baseline model includes only components which are found to have significances above $5\sigma$.
The significance is determined either based on LL difference between fits including or not the component if the added resonance was not observed before, or using ensembles of pseudoexperiments to take into account ``look-elsewhere effects'' if not~\cite{Gross10,Algeri16}.

\begin{figure*}[!tb]
\centering
\includegraphics[width=\textwidth]{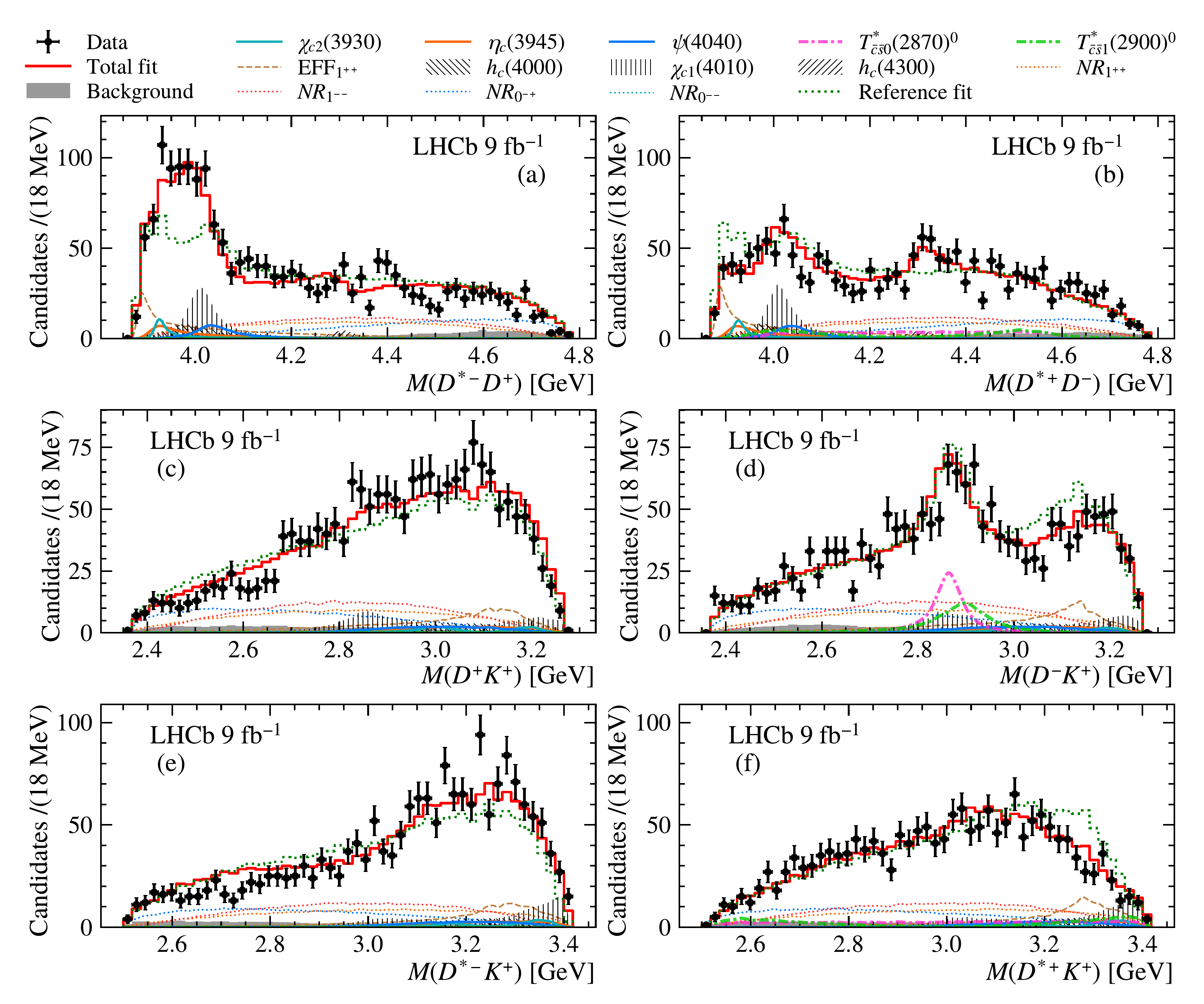}

\caption{
    Distributions of two-body invariant masses: (a)~$M(D^{*-} D^{+})$, (c)~$M(D^{+} K^{+})$ and (e)~$M(D^{*-} K^{+})$ in the $B^{+}\rightarrow D^{*-}D^{+}K^{+}$ sample; (b)~$M(D^{*+} D^{-})$, (d)~$M(D^{-} K^{+})$ and (e)~$M(D^{*+} K^{+})$ in the $B^{+}\to D^{*+}D^{-}K^{+}$ sample. 
    The fit results (red-solid lines) are overlaid on the data distributions.  
    Contributions from different components are also shown in different line styles as indicated in the legend. 
    The result of fitting the data using a model without the $h_{c}(4000)$, $\chi_{c1}(4010)$ and $h_{c}(4300)$ components (reference fit) is shown with green-dotted lines for comparison.
}\label{fig:baseline_comparison5}
\end{figure*}

\begin{table*}[!tb]
\centering
\caption{Resonant and nonresonant components included in the baseline fit and their spin parities, fit fractions and product branching fractions ($\mathcal{B}\left(B^{+}\rightarrow  R C\right)\times\mathcal{B}\left(R \rightarrow  AB\right)$), where $A, B, C$ are the three final-state particles. To obtain the branching fractions including both $R \rightarrow D^{*+}D^{-}$ and $R\rightarrow D^{*-}D^{+}$, the values in the table should be multiplied by a factor of two. The first uncertainties are statistical, estimated with a bootstrap method~\cite{efron:1979}, the second are systematic and and the third are from the uncertainty of the $B^{+}\rightarrow D^{*+}D^{-} K^{+}$ branching fraction. 
The masses and widths of the resonances marked with the ${}^{\dagger}$ symbol are fixed to their PDG values~\cite{PDG2022}.
}\label{tab:baseline_component}
\renewcommand{\arraystretch}{1.2}
\begin{tabular}{cccccccc}\hline
 Component  & $J^{P(C)}$ & \makecell{Fit fraction $[\%]$ \\ $\Bp\rightarrow\Dstarp\Dm\Kp$} & 
 \makecell{Fit fraction $[\%]$ \\$\Bp\rightarrow\Dstarm\Dp\Kp$} & \makecell{Branching fraction\\ $[10^{-4}]$ 
 }  \\\hline
 $\phantom{1} \textrm{EFF}_{1^{++}}$  & $1^{++}$  & $10.9\,_{-1.2}^{+2.3}{}\,_{-2.1}^{+1.6}$ & $9.9\,_{-1.0}^{+ 2.1}{}\,_{-1.9}^{+ 1.4}$ & $ 0.74\,_{-0.08}^{+ 0.16}{}\,_{-0.14}^{+ 0.11}\pm0.07$
 \\
 \Xnew{X0(3945)}  & $0^{-+}$  & $3.4\,_{-1.0}^{+0.5}{}\,_{-0.7}^{+1.9}$ & $3.1\,_{-0.9}^{+ 0.5}{}\,_{-0.6}^{+ 1.7}$ &
 $ 0.23\,_{-0.07}^{+ 0.04}{}\,_{-0.05}^{+ 0.13}\pm0.02 $ 
 \\
 $\phantom{1}\chi_{c2}(3930)\,^{\dagger}$  & $2^{++}$  & $1.8\,_{-0.4}^{+0.5}{}\,_{-1.2}^{+0.6}$ & $1.7\,_{-0.4}^{+ 0.5}{}\,_{-1.1}^{+ 0.6}$  & $ 0.12\,_{-0.03}^{+ 0.03}{}\,_{-0.08}^{+ 0.04}\pm0.01 $
 \\
 \Xnew{X1(4000)}  & $1^{+-}$  & $5.1\,_{-0.8}^{+1.0}{}\,_{-0.8}^{+1.5}$ & $4.6\,_{-0.7}^{+ 0.9}{}\,_{-0.7}^{+ 1.4}$ & $ 0.35\,_{-0.05}^{+ 0.07}{}\,_{-0.05}^{+ 0.10}\pm0.03 $
 \\
 \Xnew{X1(4010)}  & $1^{++}$  & $10.1\,_{-0.9}^{+1.6}{}\,_{-1.6}^{+1.3}$ & $9.1\,_{-0.8}^{+ 1.4}{}\,_{-1.4}^{+ 1.2}$ & $ 0.69\,_{-0.06}^{+ 0.11}{}\,_{-0.11}^{+ 0.09}\pm0.06$
 \\
 $\phantom{1}\psi(4040)\,^{\dagger}$  & $1^{--}$  & $2.8\,_{-0.4}^{+0.5}{}\,_{-0.5}^{+0.5}$ & $2.6\,_{-0.4}^{+ 0.5}{}\,_{-0.5}^{+ 0.4}$ & $ 0.19\,_{-0.03}^{+ 0.04}{}\,_{-0.03}^{+ 0.03}\pm0.02 $
 \\
 \Xnew{X1(4300)} & $1^{+-}$  & $1.2\,_{-0.5}^{+0.2}{}\,_{-0.2}^{+0.2}$ & $1.1\,_{-0.5}^{+ 0.2}{}\,_{-0.2}^{+ 0.2}$  & $ 0.08\,_{-0.03}^{+ 0.01}{}\,_{-0.01}^{+ 0.02}\pm0.01 $
 \\
 \hline
 $\phantom{1}\Xnew{X0(2900)}\,^{\dagger}$  & $0^{+}$  & $6.5\,_{-1.2}^{+0.9}{}\,_{-1.6}^{+1.3}$ & -- & $ 0.45\,_{-0.08}^{+ 0.06}{}\,_{-0.10}^{+ 0.09}\pm0.04 $
 \\
 $\phantom{1}\Xnew{X1(2900)}\,^{\dagger}$  & $1^{-}$  & $5.5\,_{-1.5}^{+1.1}{}\,_{-1.6}^{+2.4}$ & -- & $ 0.38\,_{-0.10}^{+ 0.07}{}\,_{-0.11}^{+ 0.16}\pm0.03 $
 \\
 \hline
 $\text{NR}_{1^{--}}(\Dstarmp\Dpm)$ & $1^{--}$ & $20.4\,_{-0.6}^{+2.3}{}\,_{-2.6}^{+2.1}$ & $18.5\,_{-0.5}^{+ 2.1}{}\,_{-2.3}^{+ 1.9}$  & $ 1.39\,_{-0.04}^{+ 0.16}{}\,_{-0.17}^{+ 0.14}\pm0.12 $
 \\
 $\text{NR}_{0^{--}}(\Dstarmp\Dpm)$  & $0^{--}$ & $1.2\,_{-0.1}^{+0.6}{}\,_{-0.6}^{+0.7}$ & $1.1\,_{-0.1}^{+ 0.6}{}\,_{-0.5}^{+ 0.6}$  & $ 0.08\,_{-0.01}^{+ 0.04}{}\,_{-0.04}^{+ 0.05}\pm0.01 $
 \\
 $\text{NR}_{1^{++}}(\Dstarmp\Dpm)$  & $1^{++}$ & $17.8\,_{-1.4}^{+1.9}{}\,_{-2.6}^{+3.6}$ & $16.1\,_{-1.3}^{+ 1.7}{}\,_{-2.3}^{+ 3.3}$ & $ 1.21\,_{-0.10}^{+ 0.13}{}\,_{-0.17}^{+ 0.24}\pm0.11$
 \\
 $\text{NR}_{0^{-+}}(\Dstarmp\Dpm)$ & $0^{-+}$  &$15.9\,_{-1.2}^{+3.3}{}\,_{-3.3}^{+3.3}$ & $14.5\,_{-1.1}^{+ 3.0}{}\,_{-3.0}^{+ 3.0}$ & $ 1.09\,_{-0.08}^{+ 0.23}{}\,_{-0.23}^{+ 0.22}\pm0.09 $
 \\
\hline
\end{tabular}
\end{table*}

The outcome of the fit with the baseline model is compared to data in Fig.~\ref{fig:baseline_comparison5}, where clear differences are seen in the $M(\Dstarp \Dm)$  and  $M(\Dstarm \Dp)$ spectra around $4.0 \gev$ due to different interference behaviours.
Numerical results and a list of the resonances included in the model are given in Table~\ref{tab:baseline_component}.
The parameters of the $\chi_{c2}(3930)$, $\psi(4040)$, $\Xnew{X0(2900)}$ and $\Xnew{X1(2900)}$ resonances are fixed to their known values~\cite{PDG2022}, while those of other states are allowed to vary freely in the fit.

At the $\Dstarpm \Dmp$ threshold, a $1^{++}$ contribution is needed to describe the spectrum. 
Different effective models ($\textrm{EFF}_{1^{++}}$) are tried including exponential NR line shapes and tails of a $\chi_{c1}(3872)$ component modelled by Eq.~\eqref{eq:SDRBW} with mass and width parameters fixed to the known values~\cite{PDG2022}. 
These two line shapes give similar fit qualities.
A model-independent partial wave approach is also attempted, which results in a similar line shape but with large uncertainties due to the increased number of parameters.  
The shape based on the $\chi_{c1}(3872)$ tail is used in the baseline model. 
It should be stressed that this is considered as an effective model, rather than a description of genuine off-shell $\chi_{c1}(3872)\to\Dstarpm \Dmp$ decays.
A fit in which the mass and width parameters in Eq.~\eqref{eq:SDRBW} are free to vary fails due to very large interference with other $1^{++}$ states. 
Moreover, the measured $\Bp \to \textrm{EFF}_{1^{++}} \Kp$, $\textrm{EFF}_{1^{++}} \to \Dstarpm \Dmp$ branching fraction is determined to be $(1.48\,^{+0.41}_{-0.35})\times 10^{-4}$, where all uncertainties are summed in quadrature. 
This value is larger than the measured branching fraction of $\Bp \to \chi_{c1}(3872) \Kp$, $\chi_{c1}(3872) \to {\Dstarz \Dzb, \Dstarzb \Dz}$~\cite{Belle:2008fma}, $(0.80 \pm 0.23)\times 10^{-4}$. 
Considering the smaller phase space available for $\chi_{c1}(3872) \to \Dstarpm \Dmp$ decays, this indicates the contribution has a more complicated nature.

Four NR contributions are included to describe the $M(\Dstarpm \Dmp)$ spectrum. 
The $\text{NR}$ line shapes are $f_{R}(m) = 1$ except for $\text{NR}_{0^{-+}}$, which is described by $f_{R}(m) = e^{(\alpha+\beta i)(m^2-m_0^2)}$ with \mbox{$m_0 = 4.35\gev$}. 
The parameters $\alpha$ and $\beta$ are determined from the fit to data to be \mbox{$0.11 \pm 0.03 \gev^{-2}$} and $-0.34 \pm 0.05 \gev^{-2}$, respectively, where the uncertainty is statistical only.   
For the $NR_{1^{++}}$ contribution, only the $S$-wave component is considered. 
The NR contributions amount to about 50\% of the total fit fraction. 
Alternative models with fewer NR contributions, or NR amplitudes in the one or more of the $D^{(*)\pm}\Kpm$ systems, are attempted but none provide agreement with the data comparable or better than that of the baseline model.

\begin{table}[!tb]
\centering
\caption{
    Comparison of the $T_{\bar{c}\bar{s}0,1}^{*0}$ properties obtained in this work to those found previously in $B^{+}\rightarrow D^{+}D^{-}K^{+}$ decays~\cite{LHCb-PAPER-2020-025}.
    In the branching fractions determined in this work, the $T_{\bar{c}\bar{s}0,1}^{*0}$ masses and widths are fixed to the previously measured values~\cite{LHCb-PAPER-2020-025}.
}
\label{tab:Xstate}
\renewcommand{\arraystretch}{1.15}
\begin{tabular}{ccc} \hline
Property   &  This work & Previous work \\
\hline 
$\Xnew{X0(2900)}$ mass [MeV] & $2914 \pm 11 \pm 15$ & $2866 \pm 7$ \\
$\Xnew{X0(2900)}$ width [MeV] & $128 \pm 22 \pm 23$  & $57 \pm 13$  \\
$\Xnew{X1(2900)}$ mass [MeV] & $2887 \pm 8 \pm 6$ & $2904 \pm 5$ \\
$\Xnew{X1(2900)}$ width [MeV] & $92 \pm 16 \pm 16$  & $110 \pm 12$ \\
\hline
$\mathcal{B}(B^+ \to \Xnew{X0(2900)} D^{(*)+})$ & $(4.5\,_{-0.8}^{+0.6}\,_{-1.0}^{+0.9}\pm 0.4)\times 10^{-5}$ & $(1.2\pm0.5)\times 10^{-5}$\\
$\mathcal{B}(B^+ \to \Xnew{X1(2900)} D^{(*)+})$ & $(3.8\,_{-1.0}^{+0.7}\,_{-1.1}^{+1.6}\pm 0.3)\times 10^{-5}$ &$(6.7\pm 2.3)\times 10^{-5}$ \\ 
\large{$\frac{\mathcal{B}(B^+ \to \Xnew{X0(2900)} D^{(*)+}) }{\mathcal{B}(B^+ \to \Xnew{X1(2900)} D^{(*)+})} $} & $1.17 \pm 0.31 \pm 0.48$ & $0.18 \pm 0.05$ \\[0.8ex]
\hline
\end{tabular}
\end{table}

The two resonant contributions, $\Xnew{X0(2900)}$ and $\Xnew{X1(2900)}$, found in $\Bp \to \Dp \Dm \Kp$ decays, are included in the $B^+ \to \Dstarp \Dm \Kp$ model to describe the enhancement seen in Fig.~\ref{fig:baseline_comparison5}(e). 
The statistical significances of the $\Xnew{X0(2900)}$ and $\Xnew{X1(2900)}$ states are found to be $11\sigma$ and $9.2\sigma$, respectively, thus confirming their existence in a new decay channel. 
If their parameters are left free in the fit, 
their values show some tension with the previous measurements, at the level of about $2\sigma$ when accounting for correlations, as seen in Table~\ref{tab:Xstate}.
In addition, the ratio of the $\Xnew{X0(2900)}$ and $\Xnew{X1(2900)}$ branching fractions in this analysis is considerably larger than in the previous work.
These tensions in the $\Xnew{X01(2900)}$ properties between $B^+ \to \Xnew{X01(2900)}\Dp$ and $B^+ \to \Xnew{X01(2900)}\Dstarp$ decays may give further hints on the $\Xnew{X01(2900)}$ production mechanism. 

The decay $\Xnew{X0(2900)} \to \Dstarm \Kp$ is forbidden by spin-parity conservation, while no clear contribution from $\Xnew{X1(2900)}$ is seen in Fig.~\ref{fig:baseline_comparison5}(c).
An upper limit on the fit fraction of $\Bp \to \Xnew{X1(2900)}D^+, \Xnew{X1(2900)} \to \Dstarm \Kp$ of 1.5\% is set at 95\% confidence level with statistical uncertainty only.
This corresponds to an upper limit on \mbox{${\cal B}(\Xnew{X1(2900)} \to \Dstarm \Kp)/{\cal B}(\Xnew{X1(2900)} \to \Dm \Kp)$} of 0.21 at 95\% confidence level, using known values of the $\Bp \to \Dp \Dm \Kp$ and \mbox{$\Bp \to \Dstarm \Dp \Kp$} branching fractions~\cite{PDG2022,LHCb-PAPER-2020-006}. Similarly, an upper limit on the fit fraction of \mbox{$\Bp \to \Xnew{Tcsbar(2900)++}D^{*-},\Xnew{Tcsbar(2900)++}\to D^{+}K^{+}$} of 3.3\% is set at 95\% confidence level.

\begin{table}[!tb]
\centering
\caption{
    Comparison of the \mbox{charmonium(-like)} states found in this analysis with previously known states and the expected $c\bar{c}$ charmonium states with relevant $J^{PC}$ quantum numbers as predicted in Ref.~\cite{Barnes05}. 
    Units of MeV for masses and widths are implied.
}\label{tab:resonances_in_th}
\resizebox{\textwidth}{!}{
\renewcommand{\arraystretch}{1.15}
\setlength{\tabcolsep}{2pt}
\begin{tabular}{cc|cc|cc}\hline
\multicolumn{2}{c|}{This work} & \multicolumn{2}{c|}{Known states~\cite{PDG2022}} &  \multicolumn{2}{c}{$c\bar{c}$ prediction \cite{Barnes05}} \\
\hline 
\Xnew{X0(3945)} & $J^{PC}=0^{-+}$& $X(3940)$~\cite{Belle:2005lik,Belle:2007woe}& $J^{PC}={?}^{??}$ & $\eta_{c}(3S)$ & $J^{PC}=0^{-+}$\\
$m_0=3945\,_{-17}^{+28}{}\,_{-28}^{+37}$ & $\Gamma_0=130\,_{-49}^{+92}{}\,_{-70}^{+101}$ & $m_0=3942\pm9$& $\Gamma_0=37\,_{-17}^{+27}$  & $m_0=4064$& $\Gamma_0=80$ \\
\hline
\Xnew{X1(4000)} & $J^{PC}=1^{+-}$ & \Xnew{X(4020)}~\cite{BESIII:2015tix}& $J^{PC}=?^{?-}$  & $h_{c}(2P)$ & $J^{PC}=1^{+-}$\\
$m_0=4000\,_{-14}^{+17}{}\,_{-22}^{+29}$ & $\Gamma_0=184\,_{-45}^{+71}{}\,_{-61}^{+97}$ & $m_0=4025.5\,_{-4.7}^{+2.0}\pm3.1$& $\Gamma_0=23.0\pm6.0\pm1.0$  & $m_0=3956$& $\Gamma_0=87$  \\
\Xnew{X1(4010)} & $J^{PC}=1^{++}$ & & & $\chi_{c1}(2P)$ & $J^{PC}=1^{++}$\\
$m_0=4012.5\,_{-3.9}^{+3.6}{}\,_{-3.7}^{+4.1}$ & $\Gamma_0=62.7\,_{-6.4}^{+7.0}{}\,_{-6.6}^{+6.4}$ & &  & $m_0=3953$& $\Gamma_0=165$\\
\hline 
\Xnew{X1(4300)} & $J^{PC}=1^{+-}$ & &  & $h_{c}(3P)$ &$J^{PC}=1^{+-}$\\
$m_0=4307.3\,_{-6.6}^{+6.4}{}\,_{-4.1}^{+3.3}$ & $\Gamma_0=58\,_{-16}^{+28}{}\,_{-25}^{+28}$  & & & $m_0=4318$& $\Gamma_0=75$ \\
& & $\chi_{c}(4274)$~\cite{LHCb-PAPER-2020-044} & $J^{PC}=1^{++}$ & $\chi_{c1}(3P)$ &$J^{PC}=1^{++}$\\
& & $m_0=4294\pm4\,_{-3}^{+6}$& $\Gamma_0=53\pm5\pm5$  & $m_0=4317$& $\Gamma_0=39$ \\
\hline
\end{tabular}
}
\end{table}
In addition to the contributions discussed above, four extra charmonium(-like) resonances are needed to describe the spectrum: $\Xnew{X0(3945)}$, $\Xnew{X1(4000)}$, $\Xnew{X1(4010)}$ and $\Xnew{X1(4300)}$, with statistical significances found to be 
$10\sigma$, $9.1\sigma$, $16\sigma$ and $6.4\sigma$, respectively.
When considering systematic uncertainties, the significance for the least significant, $\Xnew{X1(4300)}$, is $6.1\sigma$.
Their quantum numbers $J^{PC}$ are determined to be $0^{-+}$, $1^{+-}$, $1^{++}$ and $1^{+-}$, respectively, with alternative $J^{PC}$ values rejected with statistical  significances of more than 5.7$\sigma$, while other measured properties are summarized in Table~\ref{tab:resonances_in_th}. 
The assigned symbols are those for $I=0$ states~\cite{naming}, as expected for charmonium resonances produced in $\Bp \to \Dstarpm \Dmp \Kp$ decays, but the isospin quantum number is not measured and exotic contributions are possible~\cite{Dong:2021juy, Giron:2021sla}.
The mass and width of the $\Xnew{X0(3945)}$ resonance agree reasonably well with those of the previously reported $X(3940)$ state~\cite{Belle:2005lik,Belle:2007woe}.
Given the measured quantum numbers, the state could be the $\eta_c(3S)$ state predicted in Ref.~\cite{Barnes05}. 

The fit results without the $\Xnew{X1(4000)}$, $\Xnew{X1(4010)}$ and $\Xnew{X1(4300)}$ components are shown in Fig.~\ref{fig:baseline_comparison5} as green dashed lines. The $\Xnew{X1(4000)}$ and $\Xnew{X1(4010)}$ states are required to describe the discrepancy in $M(D^*D)$ around $4.0 \gev$ while the $\Xnew{X1(4300)}$ component is needed for the discrepancy around $4.3 \gev$. 
The $\Xnew{X1(4000)}$ state has $C=-1$, which generates a distinctive interference pattern with the $1^{++}$ contributions, while the $\Xnew{X1(4010)}$ component is required to describe the remaining discrepancy in this region (see Figs.~\ref{fig:diff_DstD} and~\ref{fig:contribution} in supplemental material). 
The $\Xnew{X1(4000)}$ and $\Xnew{X1(4300)}$ resonances are potential candidates for the $h_{c}(2P)$  and $h_{c}(3P)$ states, respectively, and are the first reported candidates for these two charmonium states.
The $\Xnew{X1(4000)}$ width is much larger than that of the charged $\Xnew{X(4020)}$ state found by the BESIII collaboration~\cite{BESIII:2015tix}. 
The $\chi_c(4274)$ state reported in Ref.~\cite{LHCb-PAPER-2020-044}, has mass and width close to those of the $\Xnew{X1(4300)}$ resonance but with different $C$-parity.
These are therefore likely to correspond to different charmonium states. 

The $\Xnew{X1(4010)}$ resonance has the expected $J^{PC}$ quantum numbers of a  $\chi_{c1}$ state.
Its mass is, however, larger than those of the $\chi_{c1}(3872)$ and $\chi_{c2}(3930)$ states, and smaller than those of the $\chi_{c1}(4140)$ and $\chi_{c1}(4274)$ states. 
This may indicate that the $\Xnew{X1(4010)}$ structure has exotic contributions~\cite{Wang:2023ovj, Deng:2023mza, Li:2024pfg}. 

The agreement between the data and the baseline model is not perfect, with disagreements in two regions notable in Fig.~\ref{fig:baseline_comparison5}.
Firstly, an additional structure is seen around $4.4 \gev$ in the $M(\Dstarm\Dp)$ spectrum in Fig.~\ref{fig:baseline_comparison5}(a), but not in the $M(\Dstarp\Dm)$ distribution in Fig.~\ref{fig:baseline_comparison5}(d). 
A new state is added to the model to test the significance of this structure.
The best fit is obtained with mass $4462 \pm 13\mev$, width $67 \pm 18\mev$ and $J^{PC} = 1^{++}$, where uncertainties are statistical only. 
The significance of this contribution is $3.7\sigma$, and thus it is not included in the baseline model. 
The mass and width of this contribution are close to those of the $\Xnew{Zc(4430)}$ state, however, the charge and $C$-parity are different. 
If the properties (except charge, $C$-parity and isospin) are fixed to those of the $\Xnew{Zc(4430)}$ state, the statistical significance of the component is $3.9\sigma$, a higher significance compared to when those parameters are free in the fit since the look-elsewhere effect is not considered in this case.

Secondly, the baseline model exceeds the $M(\Dstarm\Kp)$ data points in a broad range around $2.7 \gev$ in Fig.~\ref{fig:baseline_comparison5}(c).
If an additional $\Dstarm \Kp$ resonance is added to the model to address this, the fit is unstable with the width taking very large values due to interference with the ${\rm NR}_{0^{-+}}(\Dstarmp \Dpm)$ component. 
In a fit with the mass and width fixed to $2750 \mev$ and $100 \mev$, respectively, the largest LL value improvement is around 30 units when the quantum numbers are $J^P = 1^+$.  
This indicates a deficiency with the baseline model, which is treated as a source of systematic uncertainty since it is not possible to establish whether it is due to an additional resonance or a mismodeling of one or more of the NR components.

Other previously observed states~\cite{Zhukova18}, including the $\psi(4160)$, $\chi_{c1}(4274)$, $\psi(4415)$ and $\psi(4660)$ resonances are also tested in the fit, but only improve the fit quality marginally and are therefore not included in the baseline fit model.  
Other possible contributions with free mass and width under different $J^{PC}$ assumptions are also tested, and none of them improve the fit quality significantly.

The overall goodness-of-fit is quantified using binned $\chi^2$ tests and the unbinned nearest neighbor method of Refs.~\cite{Williams:2010vh,NearestNeighbor}.
These confirm the imperfect agreement between the data and the baseline model, with some metrics falling outside the 95\% confidence level for the $\Bp\rightarrow{\Dstarm\Dp\Kp}$ mode.
Acceptable goodness-of-fit, by these metrics, is obtained when the extra components mentioned above are included in the model.
Since these potential extra components are treated as a source of systematic uncertainty, the quantitative impact of the discrepancies between data and the baseline fit model is considered to be accounted for in the results.

\begin{figure}[!tb]
     \centering 
     \includegraphics[width=0.5\textwidth]{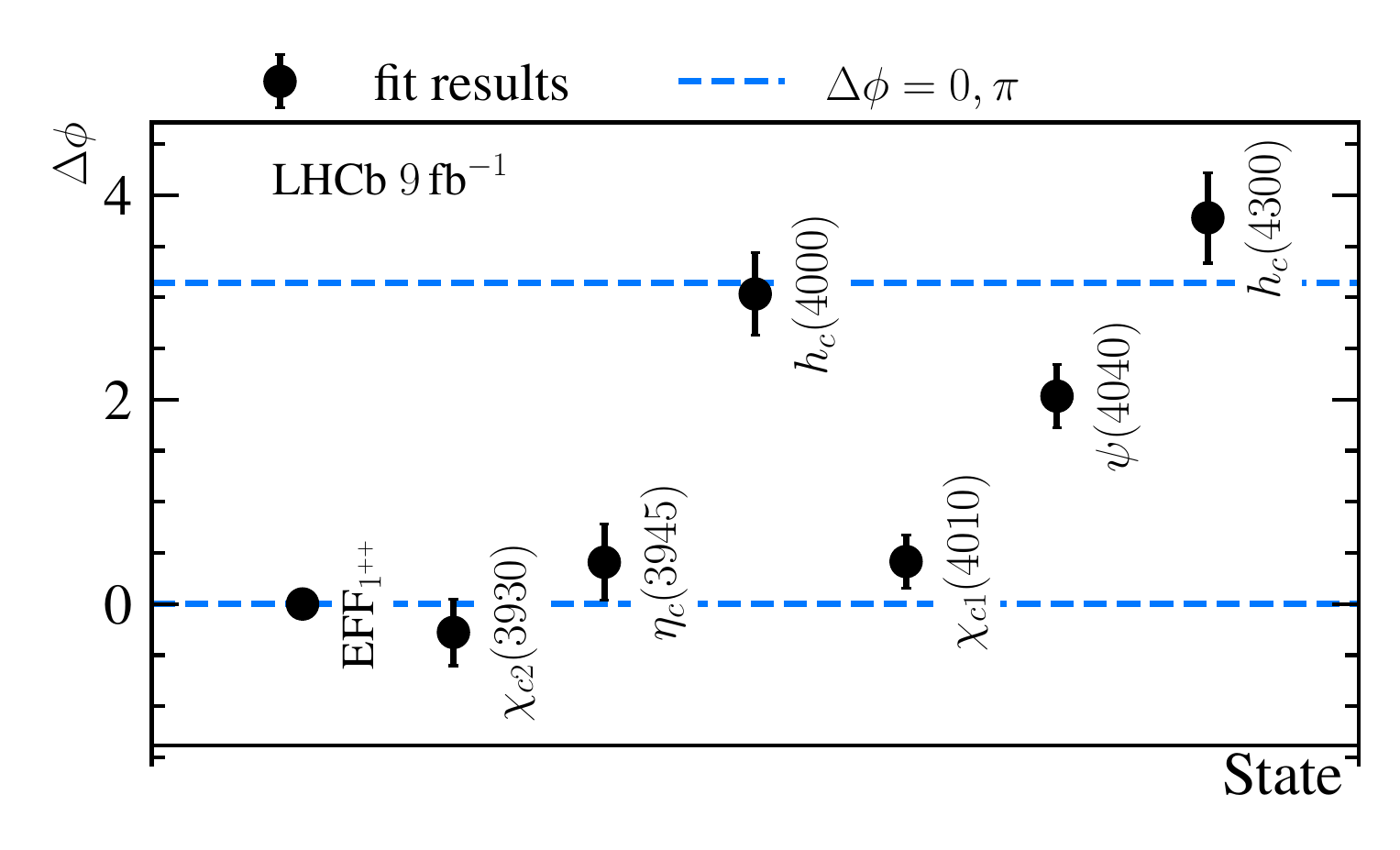}
     
     \caption{Results of the $C$-parity determination, where $C_j = \pm 1$ is replaced by $\exp(i\Delta\phi_j)$.  Uncertainties are statistical only.}
     \label{fig:float_c_parity}
\end{figure}

The main contributions in the $M(\Dstarpm\Dmp)$ distribution are states with $J^{P}=1^{+}$, which include one near-threshold contribution, three resonances and one NR component in the baseline model. 
Constructive and destructive interference effects due to the different $C$-parities of these contributions cause very different structures in Fig.~\ref{fig:baseline_comparison5}(a) and (d). 
To validate the $C$-parity relationship, a further test is performed, which replaces the value of $C$-parity in Eq.~\eqref{eq:amplitude} with $\exp(i \Delta\phi)$, where $\Delta\phi$ is a free parameter for each resonance. 
In this test, the $\textrm{EFF}_{1^{++}}$ contribution is used as a reference to set the phase convention. 
Based on Eq.~\eqref{eq:amplitude}, the phase $\phi$ should be either $\Delta\phi=0$ for $C=+$ or $\Delta\phi=\pi$ for $C=-$.  
The results are shown in Fig.~\ref{fig:float_c_parity}. 
Clear separation between $C=+$ and $C=-$ states is found, with the fit results agreeing with the $C$-parities assigned to the resonant contributions. 
The largest tension occurs for the $\psi(4040)$ state which has phase $2.7\sigma$ away from the expected value of $\pi$ when considering statistical uncertainty only. 
These results are robust against systematic variations, including the modeling of the $\textrm{EFF}_{1^{++}}$ line shape.

An alternative approach to describe the threshold enhancement is to replace the $\textrm{EFF}_{1^{++}}$ component with a $J^{PC} = 1^{+-}$ contribution, modeled by the tail of $\Xnew{Zc(3900)}$ state.
The two states lead to the same angular distributions, but their $C$-parities are different. 
The $C$-parities of other contributions, except $\chi_{c2}(3930)$ and $\psi(4040)$, are also changed. 
The alternative fit has an LL value that is worse than the baseline by 50 units, due to the unchanged $C$-parities of the well-determined $\chi_{c2}(3930)$ and $\psi(4040)$ resonances~\cite{PDG2022}.
It also results in very large NR contributions with exotic quantum numbers, which is considered unnatural. 
Thus, this hypothesis is rejected.  
Nonetheless, it is interesting to note that in this fit the new $\Xnew{X1(4300)}$ state from the baseline solution now has $J^{PC} = 1^{++}$, so that its parameters agree with those of the well-established $\chi_{c1}(4274)$ state. 
Additionally, the $\Xnew{X1(4010)}$ properties become close to those of the previously observed $\Xnew{X(4020)}$ state~\cite{BESIII:2015tix}, with some tension on the width, while the $\Xnew{X0(3945)}$ state takes exotic quantum numbers $J^{PC} = 0^{--}$.

Systematic uncertainties due to limited precision on the background fraction, background modeling, efficiency map (including simulation sample size and uncertainties due to data-simulation corrections) and fixed parameters (\ie\ resonance masses and widths) are estimated by generating pseudovalues (or distributions) according to their uncertainties. 
The root mean squares of the fit results obtained with the newly generated settings are assigned as the related systematic uncertainties.  
Systematic uncertainties due to amplitude modeling are estimated by adding other known resonances and by adding an extra resonance with freely varying mass and width under different $J^{PC}$ assumptions. 
The largest differences in fit results between all considered scenarios are assigned as the corresponding systematic uncertainties. 
The dominant systematic uncertainties on all measured quantities are those due to amplitude modeling.

In summary, a simultaneous amplitude analysis of $\Bp\rightarrow\Dstarpm\Dmp\Kp$ decays is performed for the first time.
The analysis is based on a $pp$ collision data sample collected by the \lhcb experiment, corresponding to an integrated luminosity of $9\invfb$,
and exploits
$C$-parity relations between contributions from \mbox{charmonium(-like)} resonances in the two final states. 
The results provide new insights into \mbox{charmonium(-like)} spectroscopy.
Four \mbox{charmonium(-like)} resonances, $\Xnew{X0(3945)}$, $\Xnew{X1(4000)}$, $\Xnew{X1(4010)}$ and $\Xnew{X1(4300)}$, decaying into $\Dstarpm \Dmp$, are observed, with $J^{PC}$ values determined to be $0^{-+}$,  $1^{+-}$, $1^{++}$ and $1^{+-}$, respectively. 
The $\Xnew{X0(3945)}$ resonance is found to be consistent with the previously observed $X(3940)$ state, while the other three are observed for the first time.
In addition, the results confirm the existence of the $\Xnew{X0(2900)}$ and $\Xnew{X1(2900)}$ tetraquark states in  new production channels, $B^+\to D^{*+}\Xnew{X01(2900)}$. 
The ratio of branching fractions between $B^+\to D^{*+}\Xnew{X0(2900)}$ and $B^+\to D^{*+}\Xnew{X1(2900)}$ decays is larger than the corresponding value in $B^+\to D^{+}\Xnew{X01(2900)}$ decays, which may be of interest for understanding the nature of the $\Xnew{X01(2900)}$ states. 
Studies of additional $B \to D^{(*)}\bar{D}{}^{(*)}K$ processes and new theoretical investigations will be important to provide further insight into the structures observed in this analysis.
This includes understanding the impacts of threshold effects and rescattering as well as determining the isospin of the different contributions~\cite{Bondar:2020eoa}.

\section*{Acknowledgements}
%
%
\noindent We express our gratitude to our colleagues in the CERN
accelerator departments for the excellent performance of the LHC. We
thank the technical and administrative staff at the LHCb
institutes.
We acknowledge support from CERN and from the national agencies:
CAPES, CNPq, FAPERJ and FINEP (Brazil); 
MOST and NSFC (China); 
CNRS/IN2P3 (France); 
BMBF, DFG and MPG (Germany); 
INFN (Italy); 
NWO (Netherlands); 
MNiSW and NCN (Poland); 
MCID/IFA (Romania); 
MICINN (Spain); 
SNSF and SER (Switzerland); 
NASU (Ukraine); 
STFC (United Kingdom); 
DOE NP and NSF (USA).
We acknowledge the computing resources that are provided by CERN, IN2P3
(France), KIT and DESY (Germany), INFN (Italy), SURF (Netherlands),
PIC (Spain), GridPP (United Kingdom), 
CSCS (Switzerland), IFIN-HH (Romania), CBPF (Brazil),
and Polish WLCG (Poland).
We are indebted to the communities behind the multiple open-source
software packages on which we depend.
Individual groups or members have received support from
ARC and ARDC (Australia);
Key Research Program of Frontier Sciences of CAS, CAS PIFI, CAS CCEPP, 
Fundamental Research Funds for the Central Universities, 
and Sci. \& Tech. Program of Guangzhou (China);
Minciencias (Colombia);
EPLANET, Marie Sk\l{}odowska-Curie Actions, ERC and NextGenerationEU (European Union);
A*MIDEX, ANR, IPhU and Labex P2IO, and R\'{e}gion Auvergne-Rh\^{o}ne-Alpes (France);
AvH Foundation (Germany);
ICSC (Italy); 
GVA, XuntaGal, GENCAT, Inditex, InTalent and Prog.~Atracci\'on Talento, CM (Spain);
SRC (Sweden);
the Leverhulme Trust, the Royal Society
 and UKRI (United Kingdom).




\newpage
\section*{Appendix: Supplemental material}



\begin{figure}[b]
    \centering
    \includegraphics[width=0.9\textwidth]{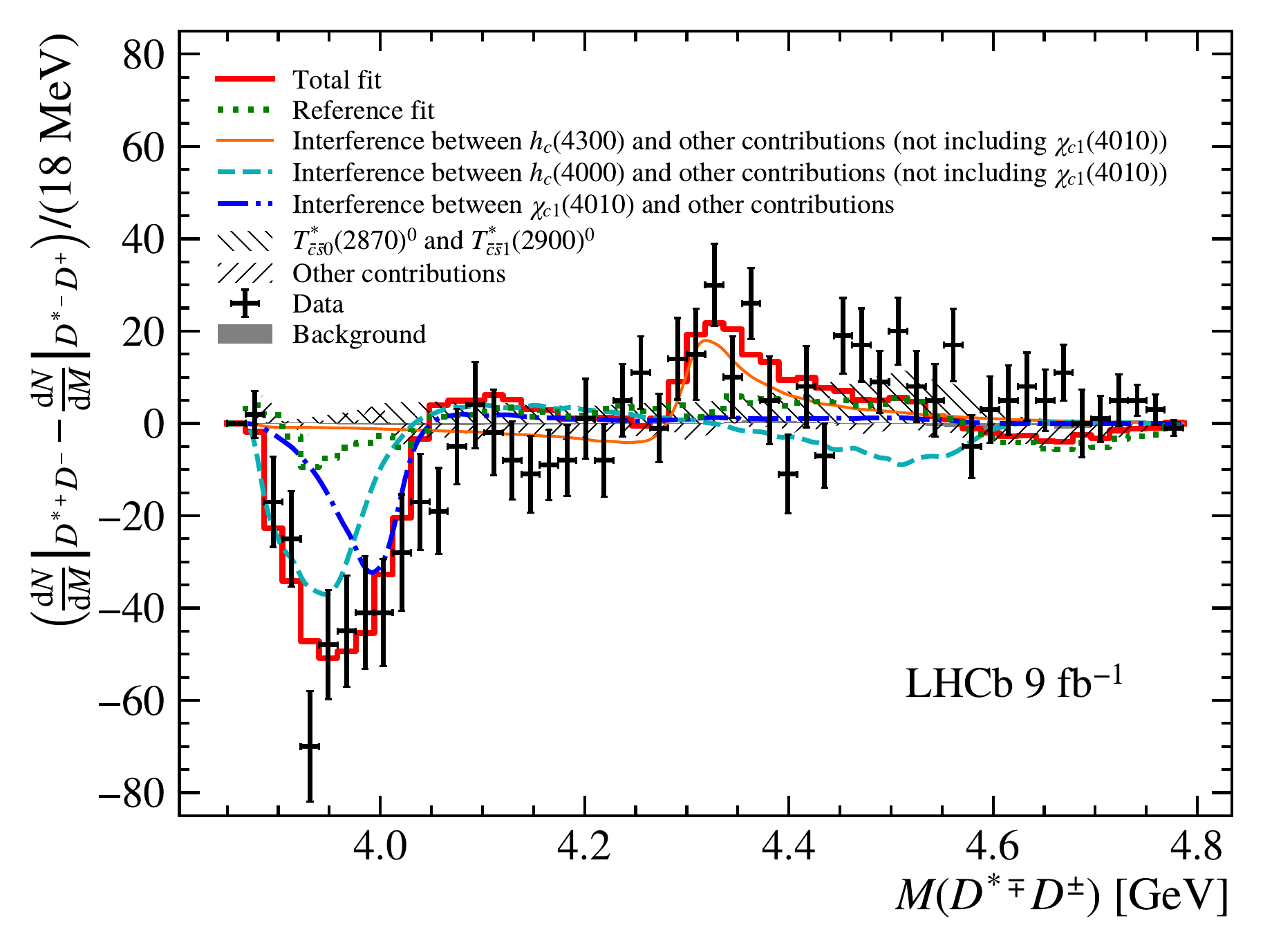}
    \caption{
    Difference between the $M(D^{*} D)$ distributions of the $B^{+}\rightarrow D^{*+}D^{-}K^{+}$ and \mbox{$B^{+}\rightarrow D^{*-}D^{+}K^{+}$} decays. Only interference between states with the same $J^P$ but different $C$-parities, and reflections from $T_{\bar{c}\bar{s}0,1}^{*0}$ resonances, have significant contributions. The reference fit where $h_c(4000)$, $\chi_{c1}(4010)$ and $h_c(4300)$ are not included is shown as green-dotted line. 
    }
    \label{fig:diff_DstD}
\end{figure}

The difference between the $M(\Dstar D)$ distributions of the $B^{+}\rightarrow D^{*+}D^{-}K^{+}$ and \mbox{$B^{+}\rightarrow D^{*-}D^{+}K^{+}$} decays is shown together with the fitted curves as in Fig.~\ref{fig:diff_DstD}. In the difference distribution, only the interference of states with the same $J^P$ but with different $C$-parties, and reflections from $\Xnew{X01(2900)}$ states, have significant contributions. Clear contributions from the three new states are visible.

\clearpage

\begin{figure}[hbp]
\centering
\includegraphics[width=0.49\textwidth]{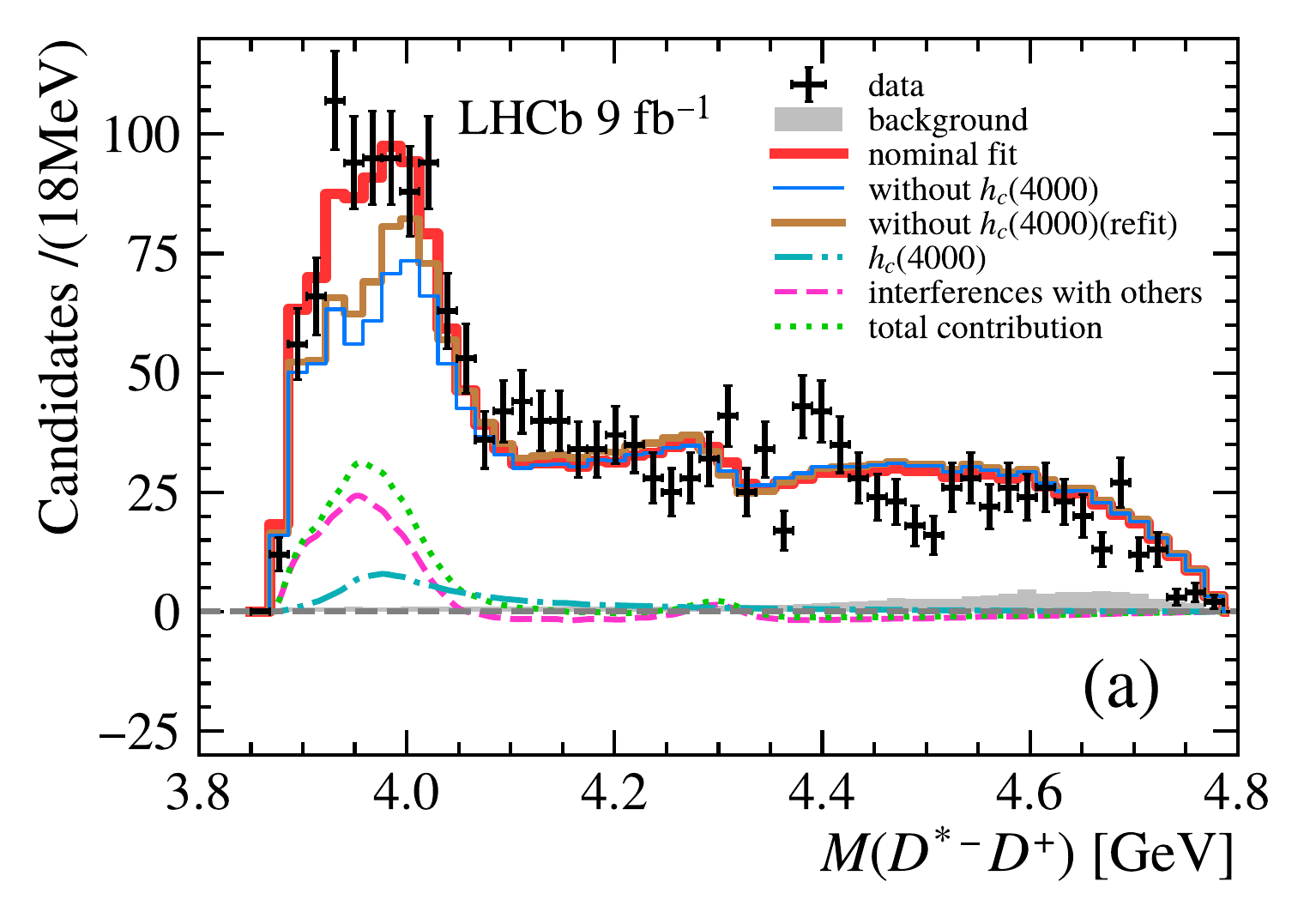}
\includegraphics[width=0.49\textwidth]{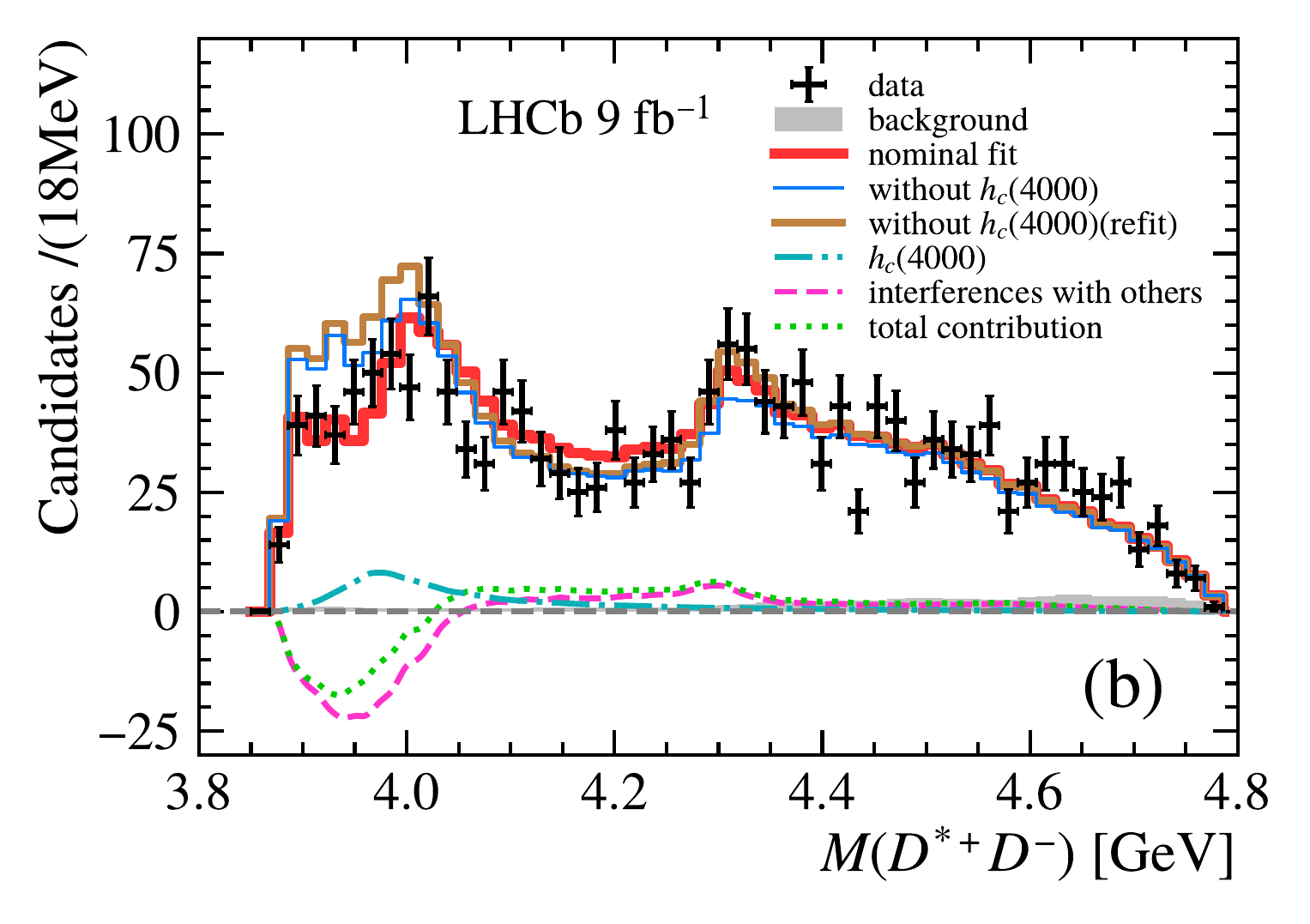}\\
\includegraphics[width=0.49\textwidth]{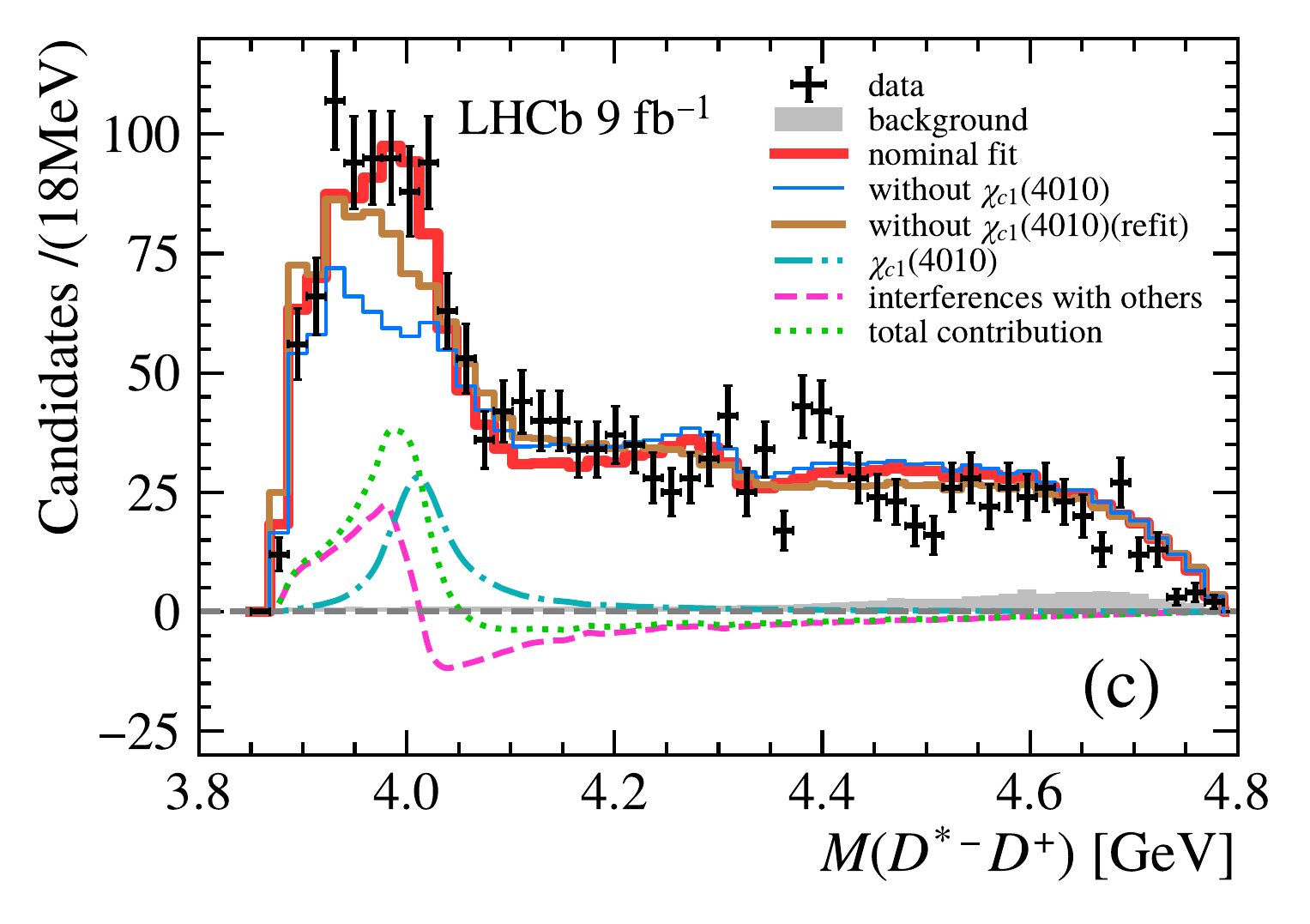}
\includegraphics[width=0.49\textwidth]{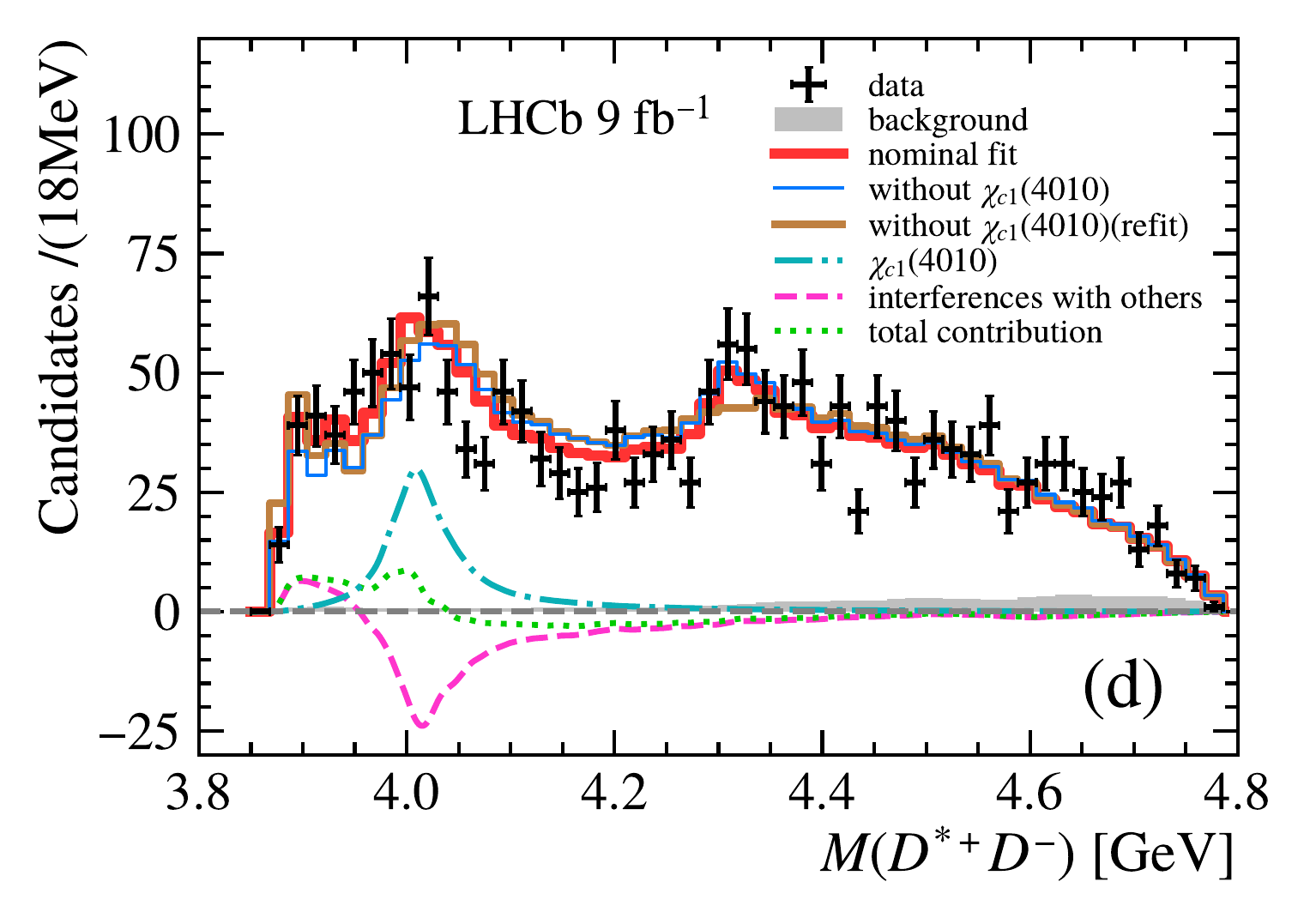}\\
\includegraphics[width=0.49\textwidth]{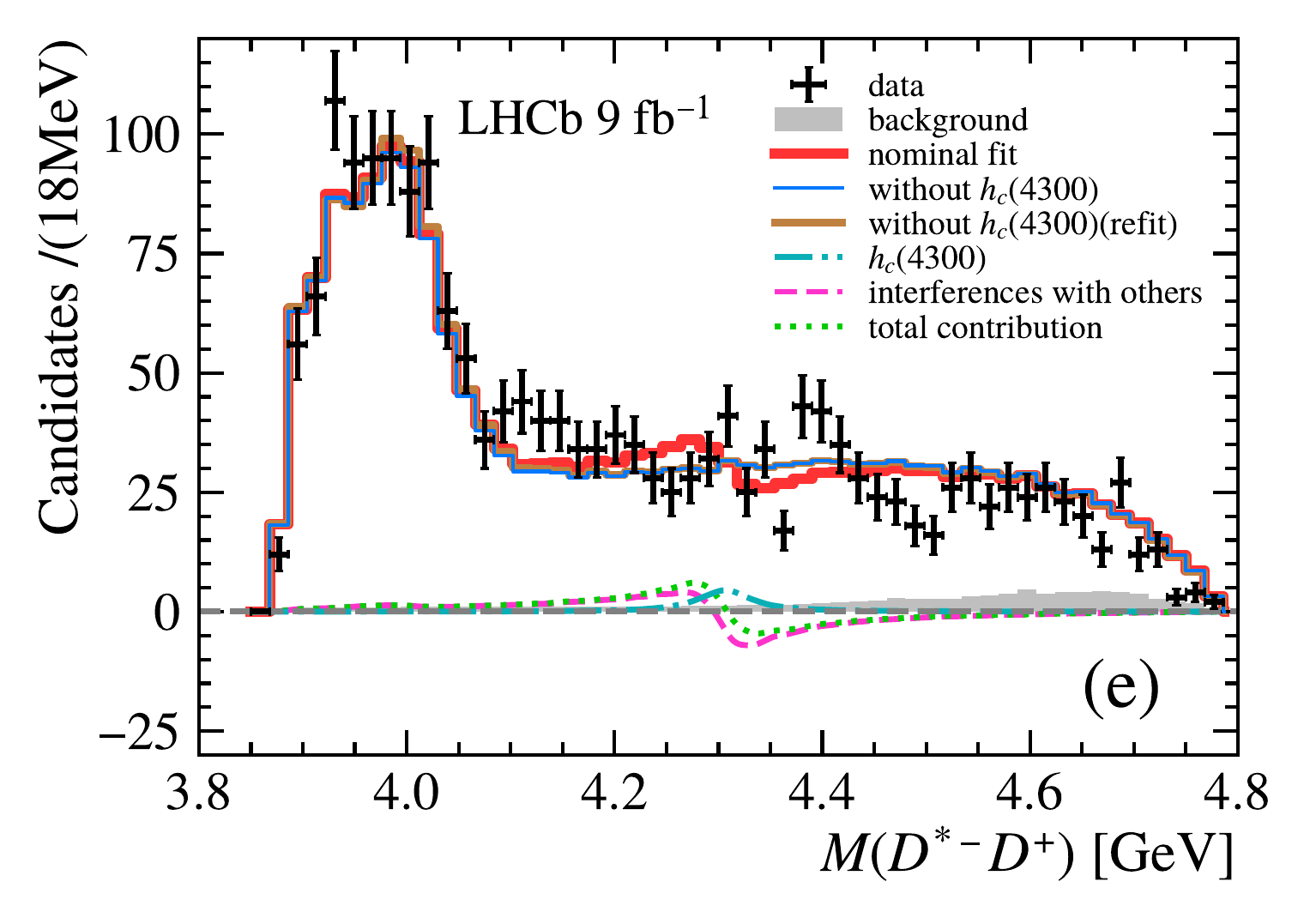}
\includegraphics[width=0.49\textwidth]{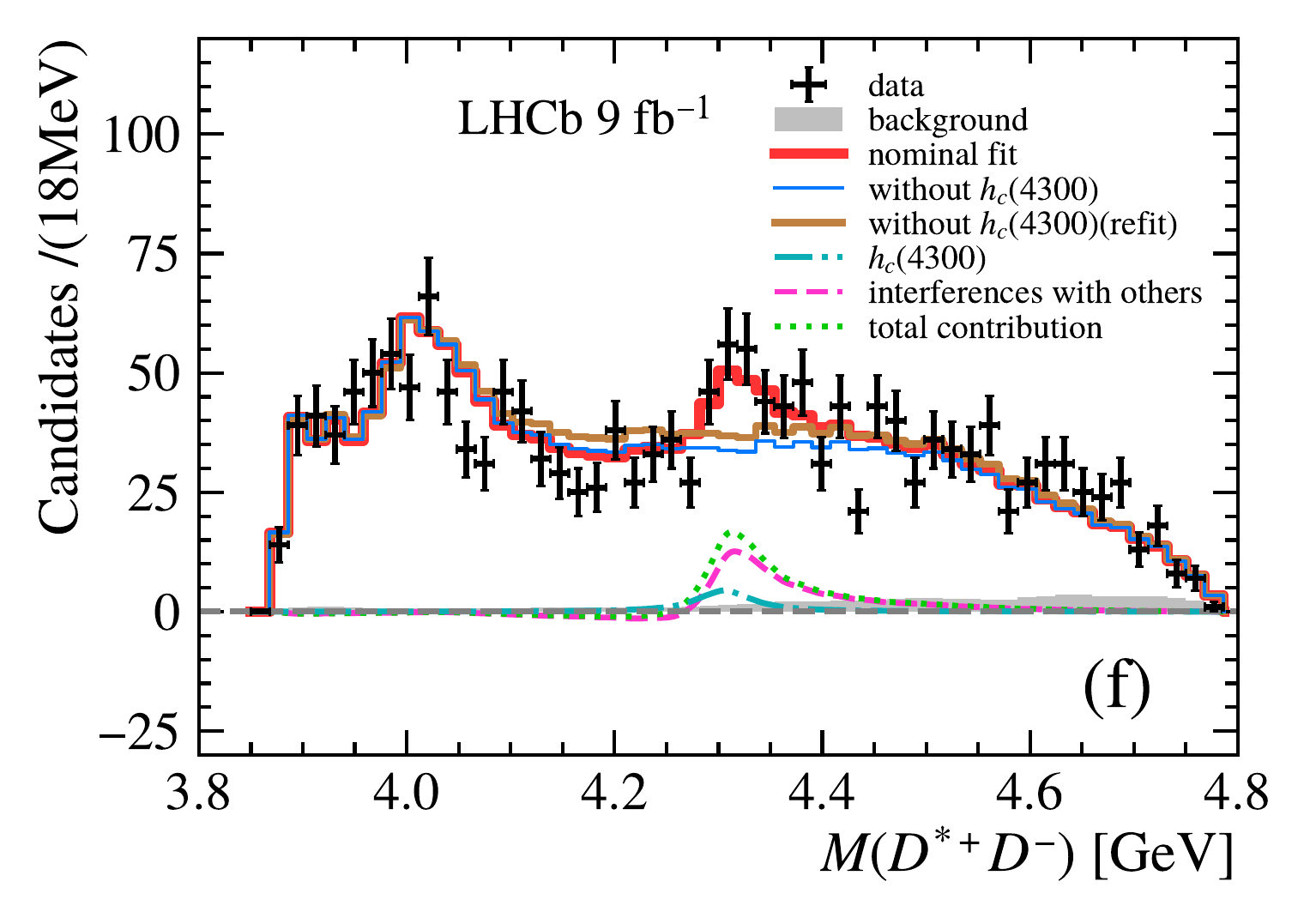}
\caption{Distributions of $M(D^{*+}D^{-})$ and $M(D^{*-}D^{+})$ with the fit results overlaid. Models with and without the new (a,b) $h_c(4000)$, (c,d) $\chi_{c1}(4010)$ and (e,f) $h_c(4300)$ states are shown in red and blue curves, respectively. The left plots (a,c,e) are from \mbox{$B^{+}\rightarrow D^{*-}D^{+}K^{+}$} decays and the right (b,d,f) from \mbox{$B^{+}\rightarrow D^{*+}D^{-}K^{+}$} decays. Contributions of (cyan) considered new states, (magenta) their interference with other contributions and (green) their total contributions are displayed on the plots. The fit model  without the considered new states (refit) are also shown as brown curves. }
\label{fig:contribution}
\end{figure}

The results of fits with and without the  $\Xnew{X1(4000)}$,$ \Xnew{X1(4010)}$ and $\Xnew{X1(4300)}$ contributions are shown in Fig.~\ref{fig:contribution}.
The fit results include the baseline model~(red) and the sum of all components in the baseline model other than the new state in question, from both fits with the baseline model~(blue) and a model with the new state removed~(brown).

\clearpage

\addcontentsline{toc}{section}{References}
\bibliographystyle{LHCb}
\bibliography{main,standard,LHCb-PAPER,LHCb-CONF,LHCb-DP,LHCb-TDR}

\newpage
\centerline
{\large\bf LHCb collaboration}
\begin
{flushleft}
\small
R.~Aaij$^{36}$\lhcborcid{0000-0003-0533-1952},
A.S.W.~Abdelmotteleb$^{55}$\lhcborcid{0000-0001-7905-0542},
C.~Abellan~Beteta$^{49}$,
F.~Abudin{\'e}n$^{55}$\lhcborcid{0000-0002-6737-3528},
T.~Ackernley$^{59}$\lhcborcid{0000-0002-5951-3498},
A. A. ~Adefisoye$^{67}$\lhcborcid{0000-0003-2448-1550},
B.~Adeva$^{45}$\lhcborcid{0000-0001-9756-3712},
M.~Adinolfi$^{53}$\lhcborcid{0000-0002-1326-1264},
P.~Adlarson$^{79}$\lhcborcid{0000-0001-6280-3851},
C.~Agapopoulou$^{47}$\lhcborcid{0000-0002-2368-0147}
% ...

{\footnotesize \it

% ...
$^{36}$Nikhef National Institute for Subatomic Physics, Amsterdam, Netherlands\\
% ...
$^{45}$Instituto Galego de F{\'\i}sica de Altas Enerx{\'\i}as (IGFAE), Universidade de Santiago de Compostela, Santiago de Compostela, Spain\\
% ...
$^{47}$European Organization for Nuclear Research (CERN), Geneva, Switzerland\\
% ...
$^{49}$Physik-Institut, Universit{\"a}t Z{\"u}rich, Z{\"u}rich, Switzerland\\
% ...
$^{53}$H.H. Wills Physics Laboratory, University of Bristol, Bristol, United Kingdom\\
% ...
$^{55}$Department of Physics, University of Warwick, Coventry, United Kingdom\\
% ...
$^{58}$School of Physics and Astronomy, University of Glasgow, Glasgow, United Kingdom\\
$^{59}$Oliver Lodge Laboratory, University of Liverpool, Liverpool, United Kingdom\\
% ...
$^{67}$Syracuse University, Syracuse, NY, United States\\
% ...
$^{79}$Department of Physics and Astronomy, Uppsala University, Uppsala, Sweden, associated to $^{58}$\\
% ...
}
\end{flushleft}

\end{document}